%% file: editor/editor.tex
\documentclass[graybox, envcountchap]{svmult}

\usepackage{makeidx}         
\usepackage{graphicx}        
\usepackage{algorithm}
\usepackage{booktabs}        
\usepackage{amsmath,amssymb,amsfonts,mathtools,nccmath,amsthm,float,bm}
\usepackage{algpseudocode,algorithmicx}
\usepackage{multicol}        
\usepackage[bottom]{footmisc}

\usepackage[varvw]{newtxmath}       

\makeindex             
\DeclarePairedDelimiter\floor{\lfloor}{\rfloor}
\newcounter{graybox}[section]  
\renewcommand{\thegraybox}{\thesection.\arabic{graybox}}  
\usepackage{etoolbox}
\makeatletter
\AtBeginDocument{%
  \patchcmd{\bibsection}
    {\addcontentsline{toc}{section}{\refname}}
    {}
    {}{}%
}
\makeatother

\begin{document}

\frontmatter

\mainmatter
\include{editor/author}
%



\end{document}

%% file: editor/author.tex
%
%
%

%
%
%
%
%
%
%

\title*{Sensing-Aided Near-Field Beam Tracking}
\author{Panagiotis Gavriilidis\orcidID{0000-0002-1967-5337} and\\ George C. Alexandropoulos\orcidID{0000-0002-6587-1371}}
\institute{Panagiotis Gavriilidis \at Department of Informatics and Telecommunications, National and Kapodistrian University of Athens, Panepistimiopolis Ilissia, 16122 Athens, Greece, \email{pangavr@di.uoa.gr}
\and George C. Alexandropoulos \at Department of Informatics and Telecommunications, National and Kapodistrian University of Athens, Panepistimiopolis Ilissia, 16122 Athens, Greece, \email{alexandg@di.uoa.gr}}
%
%
\maketitle

\abstract*{The interplay between large antenna apertures and high carrier frequencies in future wireless systems gives rise to near-field communications, where the curvature of spherical wavefronts renders traditional far-field beamforming models inadequate. This chapter investigates fundamental questions on near-field operation: (i) What is the maximum distance where far-field approximations remain effective for path gain prediction and beam design? (ii) What position resolution is needed for accurate near-field beam focusing? (iii) How frequently must channel state information be updated to maintain highly directive bweamforming in dynamic scenarios? We develop an analytical framework for assessing near-field beamforming gain degradation due to user coordinate mismatch and outdated position estimates. Closed-form expressions are derived for beam correlation, beam sensitivity to user movement, and the direction of fastest beamforming gain degradation. A dynamic polar coordinate grid is proposed for low complexity and adaptive near-field beam search. Furthermore, we introduce the novel concept of beam coherence time, quantifying the temporal robustness of focused beams and enabling proactive beam tracking strategies. Furthermore, the effect of microstrip losses on the preceding derivations is analyzed. Finally, extensive simulation results validate the theoretical analysis and the beam tracking method over randomly generated user trajectories, demonstrating significant gains over baseline methods.}

\abstract{
The interplay between large antenna apertures and high carrier frequencies in future wireless systems gives rise to near-field communications, where the curvature of spherical wavefronts renders traditional far-field beamforming models inadequate. This chapter addresses the following fundamental questions on near-field operation: (i) What is the maximum distance where far-field approximations remain effective for path gain prediction and beam design? (ii) What level of position resolution is needed for accurate near-field beam focusing? (iii) How frequently must channel state information be updated to maintain highly directive bweamforming in dynamic scenarios? We develop an analytical framework for assessing near-field beamforming gain degradation due to mismatches between the focusing point and the coordinates of a user. Closed-form expressions for beam correlation, beam sensitivity to user movement, and the direction of fastest beamforming gain degradation are derived. A dynamic polar coordinate grid is also proposed for low complexity and adaptive near-field beam search. Furthermore, we introduce the novel concept of beam coherence time, quantifying the temporal robustness of focused beams and enabling proactive sensing-aided beam tracking strategies. The effect of microstrip losses on the preceding derivations is also analyzed. Finally, extensive simulation results validate the presented theoretical analysis and beam tracking method over randomly generated user trajectories.
}

\section{Introduction}
\label{sec: Introduction}
Future generations of wireless networks are expected to incorporate high-frequency communication technologies as a key enabler for selected applications, offering access to large bandwidths that support ultra-high data rate transmissions and facilitate precise localization and sensing \cite{THz_loc_tutorial}. To mitigate the substantial path loss encountered at millimeter Wave (mmWave) and TeraHertz (THz) bands, the use of large-scale antenna arrays is being actively explored \cite{Alexandropoulos2025Measurements}. In particular, massive and holographic Multiple-Input Multiple-Output (MIMO) architectures have emerged as promising solutions \cite{Holographic_MIMO}, leveraging numerous densely placed antenna elements to produce highly directive beams capable of overcoming the substantially increased path losses in high frequencies and extend coverage \cite{XLMIMO_tutorial}.

Operating at high frequencies with eXtremely Large (XL) antenna apertures naturally brings wireless systems into the near-field propagation regime \cite{NF_tutorial}, where the curvature of wavefronts becomes significant and conventional far-field models no longer hold. In this context, spatial multiplexing can be enhanced beyond angle division multiple access through the adoption of location division multiple access schemes \cite{Hybrid_mimo_tracking,SDMA_vs_LDMA}. Additionally, the exploitation of spatial field variations enables high-precision localization and sensing using even narrowband signaling, offering an alternative to wideband techniques based on time-of-arrival estimation \cite{time_difference_arrival_localisation,RIS_aided_time_of_arrival,Localization_TOA_Primer}.

Nevertheless, scaling conventional MIMO architectures to XL apertures is constrained by hardware complexity and energy consumption. Fully digital and hybrid analog-digital solutions require numerous Radio Frequency (RF) chains and networks of phase shifters, which become impractical at large scales. Towards overcoming such shortcomings, the concept of Dynamic Metasurface Antennas (DMAs) was introduced \cite{DMA_Magazine}. A DMA is a planar array antenna architecture composed of multiple microstrips, each fed by a single RF chain and embedded with reconfigurable radiating metamaterial elements. DMAs can be considered as hybrid analog and digital MIMO architectures where their RF front end consists of a Reconfigurable Intelligent Surface (RIS) \cite{RIS_overview}, where, unlike conventional RISs, DMAs integrate RF transmit/receive chains~\cite{10352433,R-RISs}, enabling efficient hybrid beamforming with reduced hardware demands. Due to their ability to support large-scale MIMO functionality with lower RF complexity \cite{stacked_hmimo,DMA_1bit_uplink}, DMAs have attracted significant research interest in recent years \cite{gavras2023DMA_ISAC,DMA_UL_mMIMO,DMA_near_field_channel,DMA_loc_Nir,DMA_energy_eff,HMIMO_survey_et_al}. Another architecture akin to the DMA, but enabling an even greater reduction in the number of RF chains, is the two-dimensional waveguide-fed metasurface antenna \cite{pulidomancera2018}. Unlike DMAs, which rely on one-dimensional microstrip arrays, this architecture allows each RF chain to feed the entire aperture through a two-dimensional waveguide, effectively exciting a large aperture with a single RF chain. Recent work on this concept established passivity constraints for the metamaterial responses based on the energy conservation principle \cite{gavras2026waveguidefed}.

The spatial properties of the near-field regime enable not only fine-grained beamforming and spatial multiplexing, but also high precision user localization and tracking. To the best of our knowledge, \cite{CRB_source_loc} is the first work that analyzed the Cram\'{e}r-Rao Bound (CRB) limits for positioning a near-field source, while extensions like the posterior CRB that incorporated temporal dynamics for tracking applications were analyzed in \cite{NF_tracking}. A variety of architectures have been explored to exploit near-field effects, including lens-based MIMO arrays that refract impinging waves to enhance localization precision \cite{Lens_antenna_nearfield}, RIS-assisted systems employing joint optimization and Kalman filtering for dynamic tracking \cite{RIS_and_NF_tracking}, and RIS-based maximum likelihood estimators leveraging Fisher Information Matrix (FIM) analysis to jointly infer user position and channel gain \cite{RIS_localisation_George_henk_ML_estimator}. More recently, DMA-enabled localization techniques have been proposed, including approaches based on log-likelihood maximization for user positioning \cite{DMA_loc_Nir}. In addition, localization methods for DMAs with hardware imperfections, such as low esolution analog-to-digital converters \cite{Gavras2024DMALocalization1bit} and mutual coupling effects \cite{Gavras2025DMALocalizationMutualCoupling}, have also been investigated. DMAs have also been explored for Integrated Sensing and Communication (ISAC) applications \cite{ALEXANDROPOULOS2025DMA_ISAC,gavras2023DMA_ISAC,gavras2024DMA_ISAC,bayraktar2024DMA_ISAC,gavras2025DMA_ISAC}. For example, in \cite{gavras2025DMA_ISAC}, a scenario where the DMA operates as a receiver that simultaneously optimizes localization performance over an area of interest while supporting multi-user uplink communications was studied. These developments collectively highlight the growing relevance of near-field propagation for localization, tracking, and ISAC.

As aforementioned, in the near-field regime, the additional spatial degrees of freedom enable beam focusing, which produces highly directive transmission patterns capable of separating users even when they share similar angular directions. This spatial separability arises from the differences in users’ radial distances, enhancing the system’s channel multiplexing capabilities. However, the orthogonality of the resulting focusing vectors is inherently limited by both the user’s distance from the Base Station (BS) and the physical dimensions of the BS’s antenna aperture \cite{Ramezani2024}. An early study of these limitations was presented in \cite{Bjornson_dist}, which analyzed the depth of focus behavior of planar arrays for users aligned along the array’s boresight. Later works expanded this line of research by examining the spatial correlation (see, e.g.,~\cite{4786505,Peppas2011Multivariate,6184250} for the metric definition and analyses) between any two beam focusing vectors in space. For instance, \cite{Appendix_approx} derived the correlation function between two beam focusing vectors for Uniform Linear Arrays (ULAs), while \cite{SDMA_vs_LDMA} extended the analysis to Uniform Planar Arrays (UPAs) and introduced a spherical-domain codebook to be used for near-field beam searching. Moreover, the study in \cite{Near_field_NOMA} identified the conditions under which near-field beams exhibit imperfect resolution, indicating that preconfigured beams for one user can still partially serve others, thus motivating the adoption of Non-Orthogonal Multiple Access (NOMA) to exploit inter-beam correlation.

\subsection{Motivation and Contributions} \label{subsec: Motivation and Contributions}

In this chapter, the problem of sensing-aided near-field beamforming in high-frequency wireless systems is investigated, focusing on a mobile single-antenna User Equipment (UE) served by a BS equipped with a DMA transceiver. Unlike fully-connected hybrid MIMO architectures widely adopted in prior works~\cite{Hybrid_mimo_tracking,CS_DoA,Appendix_approx,vlachos2019wideband,Wideband_Hybrid_Tracking,Position_Est_mmWave}, DMAs impose a partially-connected structure with hardware-dependent analog combining constraints. As a result, existing localization and tracking algorithms, which assume fully flexible analog beamformers and unconstrained phase control~\cite{Appendix_approx,CS_DoA}, are incompatible with DMA-based architectures. We consider a scenario where the UE moves within a 2D plane, which is vertical relative to the BS with constant height offset, a common setup in practical environments~\cite{placement_of_BSs}. Motivated by the emergence of near-field communications and ultra-directive beamforming, we aim to address the following three critical questions that arise for future XL MIMO systems: \textit{i}) what is the distance limit beyond which the far-field approximation remains valid for accurate path gain prediction and beam design, \textit{ii}) how frequently should CSI be updated to maintain satisfactory beam alignment under user mobility~\cite{8313072}, and \textit{iii}) what spatial resolution is required in estimating the UE's location to preserve high-quality beam focusing performance. To this end, we introduce the notion of beam coherence time as a metric capturing the temporal stability of optimal beam configurations in Line of Sight (LoS) near-field scenarios and propose an efficient beam searching algorithm. It is noted that, although the focus in this chapter is on DMA-equipped BSs, the methodology and insights presented here are general and applicable to a broader class of planar antenna architectures.
    
This chapter's main contributions are summarized as follows:
\begin{itemize}
    \item We present an analysis of the correlation function between arbitrary beam focusing vectors in the near-field region for the considered deployment geometry. A novel analytical approximation is derived that decouples the function’s dependence on range and azimuth angle in polar coordinates. Additionally, we provide closed-form expressions for the depth of focus and beamwidth, and compare them to those obtained when accounting for signal attenuation along the microstrip line.
    \item We derive analytical expressions identifying the direction of UE movement that leads to the fastest degradation in beamforming gain. Additionally, we compute the minimum displacement required for a specified percentage of the optimal beamforming gain to be lost due to misalignment.
    \item We propose a novel non-uniform coordinate grid for adaptively sampling the searching area that the UE lies in. This grid is reconfigured dynamically at each position estimation interval and forms the basis for the proposed near-field beam tracking algorithm.
    \item We develop a sensing-aided near-field beam tracking framework in which the BS initiates a beam sweeping procedure once the estimated beamforming gain drops below a threshold relative to its optimal value. To support this process, we introduce the metric of \textit{effective beam coherence time}, defined as the minimum duration after which the UE is expected to experience a specific beamforming loss. This metric is computed online at each position estimation and guides the decision to trigger beam re-alignment.
\end{itemize}

\textit{Notations:} Vectors and matrices are represented by boldface lowercase and boldface capital letters, respectively. $\mathbf{I}_{n}$ and $\mathbf{0}_{n\times 1}$ ($n\geq2$) are the $n\times n$ identity matrix and the \(n \times 1\) zeros' vector, respectively. $[\mathbf{A}]_{i,j}$ is the $(i,j)$-th element of $\mathbf{A}$, $[\mathbf{a}]_i$ and $||\mathbf{a}||_2$ denote $\mathbf{a}$'s $i$-th element and Euclidean norm, respectively, and \(|\cdot|\) gives the amplitude (absolute value) of a complex (real) scalar. $E[\cdot]$ is the expectation operator and $\mathbf{x}\sim\mathcal{CN}(\mathbf{a},\mathbf{A})$ indicates a complex Gaussian random vector with mean $\mathbf{a}$ and covariance matrix $\mathbf{A}$, and $\jmath\triangleq\sqrt{-1}$ is the imaginary unit. Finally, \({\rm acos}(\cdot)\) is the arc cosine function. 

\section{System and Channel Models} \label{sec: System and Channel Models}
\begin{figure}[t]
    \centering
    \includegraphics[width=0.7\textwidth]{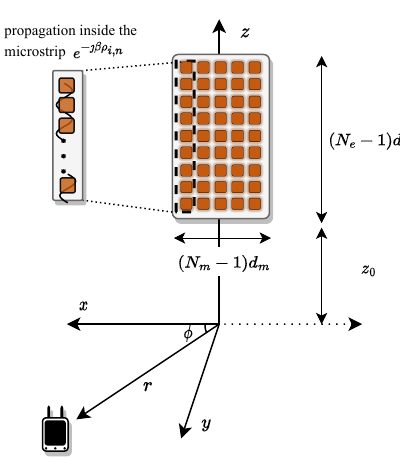}
    \caption{The considered system model comprising a static BS equipped with a DMA lying in the \(xz\)-plane and a mobile single-antenna UE moving inside the \(xy\)-plane.}
    \label{fig:System_model}
\end{figure}

Consider a point-to-point wireless communication system between a multi-antenna BS and a mobile single-antenna UE, where the BS is equipped with a DMA-based transceiver comprising \(N_m\) single-RF-fed microstrips, each including \(N_e\) metamaterial elements, making the total number of BS antenna elements \(N \triangleq N_m N_e\). In DMA architectures, it usually holds that \(N_e \gg N_m\), since scaling the number of metamaterial elements per microstrip is easier than increasing the number of RF chains~\cite{DMA_Magazine}. In this work, we concentrate on the near-field region of the BS-UE communication link, as large-aperture BS deployments are expected to operate predominantly in this regime. The Rayleigh distance, defined as \(2D^2/\lambda\), where \(D\) is the largest dimension of the aperture and \(\lambda\) denotes the signal wavelength, characterizes the transition between near-field and far-field propagation~\cite{NF_tutorial}. To define the coordinate system used in this study, let \(\phi \in [0, \pi]\) represent the azimuth angle and \(r\) the radial distance of the UE from the origin, assuming the UE moves within the \(xy\)-plane. Furthermore, the BS is assumed to lie in the \(xz\)-plane at a fixed height \(z_0\) above the UE’s plane, as illustrated in Fig.~\ref{fig:System_model}. Capitalizing on this coordinate system, we will present novel near-field beamforming designs and analysis, which would be otherwise infeasible due to the existence in the range expression formula~\cite[eq.~(10)]{SDMA_vs_LDMA} of a bilinear term, as well as the beamdepth's dependency to both the azimuth and elevation angles~\cite[eq.~(26)]{SDMA_vs_LDMA}. By referencing the coordinate system with respect to the UE's plane, the analysis is simplified. This choice is motivated by typical communication scenarios in which the UE's height coordinate rarely changes compared to the other two coordinates \cite{placement_of_BSs}.
Thereon, let \(d_e\) and \(d_m\) denote the inter-element and inter-microstrip distances, then, the Euclidean distance between each \(n\)-th metamaterial (\(n=0,1,\ldots,N_e-1\)) of each \(i\)-th microstrip (\(i=0,1,\ldots,N_m-1\)) at the BS and the UE located at the point\footnote{We henceforth make use of the compact notation \(\mathbf{p}=[r,\phi]\) for the UE position, assuming a constant height difference \(z_0\) from the BS.} \(\mathbf{p}=[r\cos(\phi),r\sin(\phi),0]\) can be expressed as follows:
\begin{align}\label{eq:Dist_approx}
    r_{i,n}\!=&\!\bigg( \!\!\left(r\cos(\phi)\!-\!i_xd_m \right)^2\!+\!\left(r\sin(\phi)\right)^2\!+\! \left(nd_e\!+\!z_0\right)^2\!\!\bigg)^{\frac{1}{2}}   \\
   & \approx\!r +\!\frac{i_x^2d_m^2}{2r}(1\!-\!\cos^{2}(\phi))\!-\!\cos(\phi)i_x d_m\!+\!\frac{n^2 d_e^2}{2 r}\!+\!\frac{z_0 n d_e}{r}\!+\!\frac{z_0}{2r},\nonumber
\end{align}
where the \((i,n)\)-th DMA element is located at the point \([i_{x} d_m, 0, nd_e+z_0]\) with \(i_{x}\triangleq i - 0.5(N_m-1)\). Therein, we have approximated \(r_{i,n}\) using the Taylor expansion \(\sqrt{1+x}\approx 1 +\frac{x}{2}-\frac{x^2}{8}\) and neglected the terms divided by \(r^2\), as per the \textit{Fresnel approximation} \cite{NF_tutorial}. However, in our coordinate system, the center of the BS is not the origin of the reference system, thus, for the maximum phase difference to be less than \(\frac{\pi}{8}\) at \(\phi = \frac{\pi}{2}\), we employ the Lagrange error bound \cite{apostol1991calculus}, which yields the following condition, so that the approximation in \eqref{eq:Dist_approx} holds: 
\begin{align}\label{eq:r_bound}
    r\geq r_{\rm appr} \triangleq \left(\frac{2L^4_{z_0}}{\lambda}\right)^{1/3},
\end{align}
where \(L_{z_0}\triangleq\sqrt{((N_e-1)d_e + z_0)^2 + 0.25(N_m-1)^2 d^2_m}\).
Equivalently, denoting as \(r_0\) the distance from the DMA's center, i.e., \(r_0 = \sqrt{r^2 +(z_0+0.5(N_e-1))^2 }\), the approximation in \eqref{eq:Dist_approx} holds for \(r_0 \geq r_{0,{\rm appr}} \triangleq \sqrt{r_{\rm appr}^2 +(z_0+0.5(N_e-1))^2} \). It is noted that this result aligns with the literature when bringing the coordinate system to the DMA's center; this is accomplished by setting \(z_0= -0.5(N_e-1)d_e\). For this origin shifting, the approximation holds for \(r_0 \geq \frac{D}{2} \left(\frac{D}{\lambda}\right)^{1/3}\)~\cite[eq.~(19)]{Rayleigh_Fresnel_distances}. It is important to emphasize that the results presented in this work retain their three-dimensional nature. We merely assume that the height difference \(z_0\) remains constant during communication; however, \(z_0\) can take any arbitrary value. As will be shown in Statement~\ref{box: Dr beamfocusing}, this value influences the derived expressions for the depth of focus.

\subsection{DMA Modeling}
The BS-UE channel vector is both difficult and time-consuming to acquire due to the extremely massive number of antennas at the BS, which is why in XL MIMO communications a spatial representation of the channel is used to effectively reduce the dimensionality of the channel estimation problem~\cite{vlachos2019wideband,Wideband_Hybrid_Tracking}. In this chapter, we focus on DownLink (DL) communications and aim to provide beam focusing toward the UE's position, that is, we wish to maximize the achievable gain to the location where the UE lies. Let $s$ denote the unit-power information symbol transmitted via the beamformed vector \(\mathbf{x} \triangleq \mathbf{\bar{Q}}\mathbf{v}s\in \mathbb{C}^{N \times 1}\),  where $\mathbf{\bar{Q}}\in \mathbb{C}^{N \times N_m}$ and \(\mathbf{v}\in \mathbb{C}^{N_m \times 1}\) represent the DMA's analog and digital beamformers, respectively.
The transmitted vector is assumed power limited as \(||\mathbf{x}||^2_2\leq P_b\) with \(P_b\) being the maximum BS transmit power. The DMA analog beamformer is defined as \({\mathbf{\bar{Q}}} \triangleq \mathbf{P}_{m}\mathbf{Q}\), where the diagonal matrix \(\mathbf{P}_m\in \mathbb{C}^{N \times N}\) models signal propagation inside each microstrip and \(\mathbf{Q} \in \mathbb{C}^{N \times N_m}\) includes the tunable responses of the identical metamaterials~\cite{DMA_near_field_channel}, which we henceforth term as analog weights. More specifically, for each \(n\)-th element of each \(i\)-th microstrip, assuming lossless propagation at each microstrip, the phase distortion due to the intra-microstrip propagation is modeled as \cite[eq.~(8)]{DMA_effective_aperture}: 
\begin{equation}\label{eq:P_matrix}
 [\mathbf{P}_m]_{iN_e+n, iN_e+n}\triangleq \exp\left(-\jmath \beta\rho_{i,n}\right),
\end{equation}
where \(\beta \triangleq 2 \pi\lambda^{-1}\sqrt{\epsilon}_r\) is the microstrip’s wavenumber with \(\epsilon_r\) being its dielectric constant, and \(\rho_{i,n}\) denotes the distance of the \(n\)-th element in the \(i\)-th microstrip from the
RF chain port. Finally, the elements of the matrix $\mathbf{Q}$, which are the analog weights, are assumed to follow a Lorentzian-constrained phase model, according to which:
\begin{equation}\label{eq:Q_matrix}
[\mathbf{Q}]_{iN_{e}+n,j}=\begin{cases} q_{i,n}\in \mathcal{Q},&i=j\\0,&i\neq j\end{cases}
\end{equation}
with $\mathcal{Q}\triangleq\left\{0.5\left(\jmath + e^{\jmath\phi}\right)/\sqrt{N_e} | \phi \in \left[0,2\pi\right]\right\}$. 

The Lorentzian-constrained phase profile is commonly used for modeling radiating dipoles~\cite[eq.~(1)]{DMA_effective_aperture}, which explains its vast application in characterizing the DMA's analog weights \cite{DMA_near_field_channel,DMA_UL_mMIMO,DMA_loc_Nir}. We note here that the normalization with respect to \(N_e\) is done so that the energy conversation principle is not invalidated during transmission \cite{gavriilidis2025_microstrip_losses}. That is, the power of the signal emanating from the RF chain attached to the microstrip line should be less than the collective power transmitted from the metamaterials in that microstrip, since there is no amplification involved but only ``passive'' metamaterials with reconfigurable transmission characteristics. On the other hand, during reception with DMAs this normalization term can be omitted, since each metamaterial captures its own signal which bares its respective power and there is no constraint on the combined power.

\subsection{Channel Model}

In this subsection, we introduce the channel model that will be used throughout this work. In the near-field regime, the primary distinction from the far-field model lies in the replacement of the conventional steering vector with a focusing vector that accounts for the distance-dependent phase variations. Consequently, the LoS channel vector \( \mathbf{h}_{\text{LoS}} \in \mathbb{C}^{N \times 1} \) between the BS and the UE is modeled as~\cite{NF_tutorial}:
\begin{align}\label{eq:LoS_TX_to_UE_channel}
    [\mathbf{h}_{\text{LoS}}]_{iN_e+n} \triangleq \frac{\lambda}{4 \pi r_{i,n}} e^{-\jmath \frac{2 \pi}{\lambda} r_{i,n}},
\end{align}
where \( r_{i,n} \) denotes the distance between the \( n \)-th element of the \( i \)-th microstrip in the DMA and the UE. To account for environmental scatterers and possible Non-Line-of-Sight (NLoS) components, the complete channel vector is expressed as:
\begin{equation}\label{eq:TX_to_UE_channel}
    \mathbf{h} \triangleq \mathbf{h}_{\text{LoS}} + \sum_{\ell=1}^{L} \mathbf{h}_{\text{NLoS},\ell},
\end{equation}
where \( [\mathbf{h}_{\text{NLoS},\ell}]_{iN_e+n} \triangleq g_{\ell} \frac{\lambda}{4 \pi r_{\ell,i,n}} e^{-\jmath \frac{2 \pi}{\lambda} r_{\ell,i,n}} \). The coefficient \( g_{\ell} \triangleq e^{-\jmath w} e^{-\jmath \frac{2\pi}{\lambda} d_{\ell}} \frac{\lambda}{4 \pi d_{\ell}} \) captures the reflection-induced attenuation and phase shift, with \( d_{\ell} \) and \( r_{\ell,i,n} \) denoting the distances from the UE to the \( \ell \)-th scatterer and from the scatterer to the \( (i,n) \)-th DMA element, respectively, based on the linear reflection model~\cite{RIS_energy_eff}. The variable \( w \sim \mathcal{U}(-\pi,\pi) \) models a uniformly distributed random phase due to reflection. In this work, we focus on the radiative near-field region, defined by \( r_{\text{FD}} \leq r_0 \leq r_{\text{RD}} \), where \( r_{\text{FD}} \triangleq 0.62 \sqrt{D^3 / \lambda} \) and \( r_{\text{RD}} \triangleq 2 D^2 / \lambda \) denote the Fresnel and Rayleigh distances, respectively~\cite{Rayleigh_Fresnel_distances}. Within this region, the pathloss across the DMA aperture can be approximated as constant~\cite{NF_tutorial}, allowing the LoS component in \eqref{eq:LoS_TX_to_UE_channel} to be reformulated as:
\[
\mathbf{h}_{\text{LoS}} = \frac{\lambda}{4 \pi r_0} \mathbf{a}(r, \phi),
\]
where the focusing vector is defined as \( [\mathbf{a}(r, \phi)]_{iN_e+n} \triangleq e^{-\jmath \frac{2 \pi}{\lambda} r_{i,n}} \).

The narrowband received signals at the UE, in DL, and at the outputs of the $N_m$ microstrips of the BS, in the UpLink (UL), are given as:
\begin{align}
     y_u\triangleq  & \mathbf{h}^{\rm H}\mathbf{x} + n_u, \label{eq:Rx_UE}\\
    \mathbf{y}_b\triangleq &\mathbf{\bar{Q}}^{\rm H}\mathbf{h}x_u + \mathbf{\bar{Q}}^{\rm H} \mathbf{n}_b, \label{eq:DMA_UE}             
\end{align}
where \(n_u \sim\mathcal{CN}(0,\sigma^2)\) and \(\mathbf{n}_b\sim\mathcal{CN}(\mathbf{0}_{N\times 1},\sigma^2\mathbf{I}_{N}) \) represent the Additive White Gaussian Noise (AWGN) contributions at the UE and BS, respectively, and \(x_u\triangleq \sqrt{P_u}\), with \(P_u\) being the UE's transmit power, denotes the pilot symbol transmitted from the UE to enable its localization at the BS side.

 \section{Near-Field Beamforming Analysis}\label{sec: BF_analysis}
In this section, the optimal DMA-based near-field beamformer is first presented under the assumption of perfectly known UE position coordinates. Subsequently, analytical expressions for the relative beamforming gain in the presence of position mismatch are derived. Finally, the \textit{effective beam coherence time} metric is introduced, which will be employed in Section~\ref{sec:Proposed_BA} as part of the proposed near-field beam tracking framework.

\subsection{Beamforming at Known UE Position}\label{sec:Optimal_BF}
The analog beamformer at the DMA that maximizes the Signal-to-Noise Ratio (SNR) at a given UE position \(\mathbf{p} = [r,\phi]\) is designed by phase-aligning with \(\mathbf{h}_{\rm LoS}^{\rm H}\), as derived in~\cite[Lemma 1]{DMA_loc_Nir}. Specifically, the analog weights are set as \(q_{i,n} = 0.5\left(\jmath + e^{\jmath(\angle{[\mathbf{a}(r,\phi)}]_{iN_e+n} + \beta \rho_{i,n})}\right)\) for all \(i,n\). When only the UE position coordinates are available, this configuration constitutes the optimal beamformer. We define the beamforming gain at position \(\mathbf{p}\) as \(G \triangleq |\mathbf{a}^{\rm H}(r,\phi)\mathbf{\bar{Q}}\mathbf{v} |^{2}\). Setting the analog beamformer \(\mathbf{\bar{Q}}\) using the weights above, and optimizing \(G\) with respect to the digital weights \(\mathbf{v}\), the optimal solution corresponds to a vector of ones scaled to satisfy the transmit power constraint~\cite[Proposition~1]{gavriilidis2025_beam_tracking}. This yields an optimal beamforming gain of \(G_{\rm opt} = 0.25\, P_b N_m N_e\)~\cite{gavriilidis2025_microstrip_losses}, while, the optimal hybrid (analog-digital) DMA configuration is:
\begin{equation}\label{eq: Hybrid Analog and Digital Config for Los focusing}
\mathbf{\bar{Q}}\mathbf{v}= 0.5\sqrt{\frac{P_b}{N}}\left(\mathbf{a}(r,\phi)  + \jmath \,e^{-\jmath \beta \boldsymbol{\rho}}\right),
\end{equation}
where the term \(e^{-\jmath \beta \boldsymbol{\rho}}\) accounts for signal propagation within the microstrips and does not affect the beamforming gain. When the beam is steered toward an incorrect position \(\mathbf{\hat{p}} \triangleq [\hat{r}, \hat{\phi}]\), the resulting hybrid beamformer is given by:
\[
\mathbf{\bar{Q}}\mathbf{v}_{\rm opt} = 0.5\sqrt{\frac{P_b}{N}}\left(\mathbf{a}(\hat{r},\hat{\phi}) + \jmath \,e^{-\jmath \beta \boldsymbol{\rho}}\right),
\]
and the corresponding beamforming gain at the true UE location \(\mathbf{p} = [r,\phi]\) becomes
\begin{equation}
G_{\hat{r},\hat{\phi}} \triangleq 0.25 \frac{P_b}{N} \left|\mathbf{a}^{\rm H}(r,\phi)\mathbf{a}(\hat{r},\hat{\phi}) \right|^2.
\end{equation}
The relative beamforming gain, also referred to as the correlation factor, quantifies the loss due to position mismatch and is defined as the ratio between the achieved and optimal gain:
\begin{equation}
\frac{G_{\hat{r},\hat{\phi}}}{G_{\rm opt}} = \frac{\left|\mathbf{a}^{\rm H}(r,\phi)\mathbf{a}(\hat{r},\hat{\phi}) \right|^{2}}{N^2}.
\end{equation}
This expression essentially quantifies the fraction of the optimum gain achieved due to misfocusing.
In what follows, we analytically characterize this correlation factor as a function of the mismatch between the actual UE position and the BS's target focusing point.

\subsection{Beamforming Gain under UE Coordinate Mismatches}\label{suubsec: BF_gain_mmismatch}
We begin by investigating the beamforming loss due to a range mismatch, i.e., when \(\hat{r}=r\pm \Delta r\) with \(\Delta r \geq 0\) \cite[Lemma~1]{gavriilidis2025_beam_tracking}.
\refstepcounter{graybox}
\label{box: Dr beamfocusing}
\begin{svgraybox}
\textbf{Statement~\thegraybox:} When the BS beam focuses at the point \((\hat{r}, \phi)\), the relative beamforming gain is obtained as~\(\frac{G_{\hat{r},\phi}}{G_{\rm opt}}\triangleq\frac{|\mathbf{a}^{\rm H}(r,\phi)\mathbf{a}(r \pm \Delta r,\phi) |^{2}}{N^2 } = \mathcal{K}^2\left(a(\pm \Delta r), \phi\right)\), where function \( \mathcal{K}\left(x, \phi\right) \triangleq I(x)\left(1-\frac{\pi^2}{90}\left(x \frac{(N_m-1) d_m}{2(N_e-1) d_e}|\sin(\phi)|\right)^4\right)\) 
with \(a(x) \triangleq \sqrt{\frac{2|x|}{r^2 + r x}}  \frac{d_e(N_e-1)}{\sqrt{\lambda}}\), and function $I(x)$ defined as \cite[Lemma~1]{gavriilidis2025_beam_tracking}:
    \begin{align}\label{eq: prop_Dr}
         I(x) \triangleq \frac{1}{x}&  \Bigg |\left[C\left(x+\frac{x z_0}{(N_e-1)d_e}\right)-C\left(\frac{x z_0}{(N_e-1)d_e}\right)\right] + \nonumber\\
        & \jmath \left[S\left(x+\frac{x z_0}{(N_e-1)d_e}\right)-S\left(\frac{x z_0}{(N_e-1)d_e}\right)\right]\Bigg | ,
    \end{align} 
    where \(C(\cdot)\) and \(S(\cdot)\) are the Fresnel functions \cite[eq. (12)]{Fresnel}. 
\end{svgraybox}
$\mathcal{K}\left(a(\pm \Delta r), \phi\right)$  has low-to-zero dependency on \(\phi\), which holds due to the fact that\footnote{{  At most cases, \(\left(a(\pm \Delta r) \frac{(N_m-1) d_m}{2(N_e-1) d_e}|\sin(\phi)|)\right)<1\) since \(N_e > N_m\), and for the considered range mismatches \(\Delta_{\kappa}^{\pm}(r)\), it typically holds that \(a(\pm\Delta_{\kappa}^{\pm}(r))< 1\).}}: \(\frac{\pi^2}{90}\left(a(\pm \Delta r) \frac{(N_m-1) d_m}{2(N_e-1) d_e}|\sin(\phi)|)\right)^4 \to 0\), implying the tight approximation: \(\mathcal{K}\left(a(\pm \Delta r), \phi\right) \approx I(a(\pm \Delta r))\). This approximation helps us to estimate the \(\pm\Delta r\) terms that result in a \((1-\kappa \%\)) beamforming gain loss, by simply setting \(I^2\left(a(\pm \Delta r)\right) = 0.01\kappa\) and solving with respect to \(a(\pm \Delta r)\). The solution to the latter equation is denoted by \(a_{\kappa}\), and if there exist more than one solution, we select the smallest positive value.
Consequently, to solve for \(\pm\Delta r\), we set \(a(\pm \Delta r)=a_{\kappa}\) and obtain the following solutions:
\begin{equation}\label{eq:Dr}
  \Delta^{\pm}_{\kappa}(r) = r^2\left(\frac{2 d_e^2 \left(N_e-1\right)^2}{\lambda a_{\kappa}^2} \mp r\right)^{-1}.
\end{equation}
These solutions can be used to determine the depth of the BS beam focusing region within which the relative beamforming gain does not fall below \(\kappa\%\), that is, \(\frac{G_{\hat{r},\phi}}{G_{\rm opt}} \geq \kappa\%\) for \(\hat{r} \in [r - \Delta^{-}_{\kappa}(r),\; r + \Delta^{+}_{\kappa}(r)]\). It should be noted that as \(r \to r_{{\rm lim},\kappa} \triangleq \frac{2 d_e^2 (N_e - 1)^2}{\lambda a_{\kappa}^2} \approx \frac{r_{\rm RD}}{a_{\kappa}^2}\), the upper bound \(\Delta^{+}_{\kappa}(r)\) diverges to infinity. This represents a limiting distance beyond which no finite \(\Delta^{+}_{\kappa}(r)\) satisfies \(\frac{|\mathbf{a}^{\rm H}(r,\phi)\mathbf{a}(\hat{r},\phi)|^{2}}{N^2} \geq \kappa\%\) for \(\hat{r} > r + \Delta^{+}_{\kappa}(r)\), as the UE has already entered the far-field and continues to move away from the BS. In contrast, the quantity \(\Delta^{-}_{\kappa}(r)\) always exists, as it corresponds to movement towards the BS, i.e., deeper into its near-field region.

Furthermore, if we incorporate microstrip propagation losses into \eqref{eq:P_matrix}, that is, \([\mathbf{P}_m]_{iN_e+n, iN_e+n}\triangleq \exp\left(-(\alpha +\jmath \beta)\rho_{i,n}\right)\), where \(\alpha\) denotes the attenuation coefficient, then the correlation function becomes attenuation-dependent, expressed as \(I(x,\alpha)\) and is formulated via the imaginary error function \({\rm erfi}(\cdot)\), as detailed in \cite[Lemma~1]{gavriilidis2025_microstrip_losses}. 
In what follows, we numerically validate the range limits \(\Delta^{\pm}_{\kappa}(r)\) and assess the accuracy of the closed-form expression in \eqref{eq: prop_Dr}, which was derived under the assumption of lossless propagation within the microstrip, by comparing it against realistic lossy scenarios using commercially available substrates \cite[Table~14.1]{balanis2016antenna}. In general, the range limits are attenuation-dependent and can be expressed as \(\Delta^{\pm}_{\kappa}(r,\alpha)\). While a closed-form expression for \(\Delta^{\pm}_{\kappa}(r,\alpha)\) is available in \cite[eq.~(11)]{gavriilidis2025_microstrip_losses}, analyzing the lossless case enables more tractable derivations and clearer insights. For comparison purposes, we consider the height \(z_0\) to be \(-0.5 (N_e-1) d_e\), since the findings in this chapter are derived for polar coordinates while the findings in \cite{gavriilidis2025_microstrip_losses} consider spherical coordinates. With this choice of \(z_0\) the azimuth angle and range in polar and spherical coordinates align.

Another key consideration in the lossy case is that the optimal gain \(G_{\rm opt}\) varies with \(\alpha\), i.e., \(G_{\rm opt} = G_{\rm opt}(\alpha)\), as given in \cite[eq.~(6)]{gavriilidis2025_microstrip_losses}. To enable consistent performance comparisons, we normalize the beamforming gain as \(\frac{G(\hat{r},\hat{\phi})}{G_{\rm opt}(\alpha)}\) for the lossy cases, and evaluate the range limits where a specified fraction of the peak gain is preserved. Based on \cite[Table~14.1]{balanis2016antenna}, we examine two mmWave-compatible substrates: i) Duroid 5880 with \(\alpha = 0.7381\,\text{m}^{-1}\), and ii) RO 3003 with \(\alpha = 0.8629\,\text{m}^{-1}\), assuming a strip width of \(\lambda/2\), height \(\lambda/20\), and operation at \(f_c=30\) GHz. In \cite{gavriilidis2025_beam_tracking}, it was shown that the impact of attenuation becomes significant when the auxiliary variable \(w \triangleq 0.5 N_e d_e \alpha\) exceeds \(3.1\), where \(w\) represents the attenuation exponent at the microstrip's midpoint. Considering the aforementioned parameters, this translates to a maximum number of elements per microstrip of i) \(N_e = 1680\) for Duroid 5880 and ii) \(N_e = 1432\) for RO 3003. Therefore, for moderate microstrip lengths, the lossless formula in \eqref{eq: prop_Dr} remains accurate even when attenuation is present \cite[eq.~(11)]{gavriilidis2025_microstrip_losses}.

\begin{figure}[t]
    \centering
    \includegraphics[width=1\textwidth]{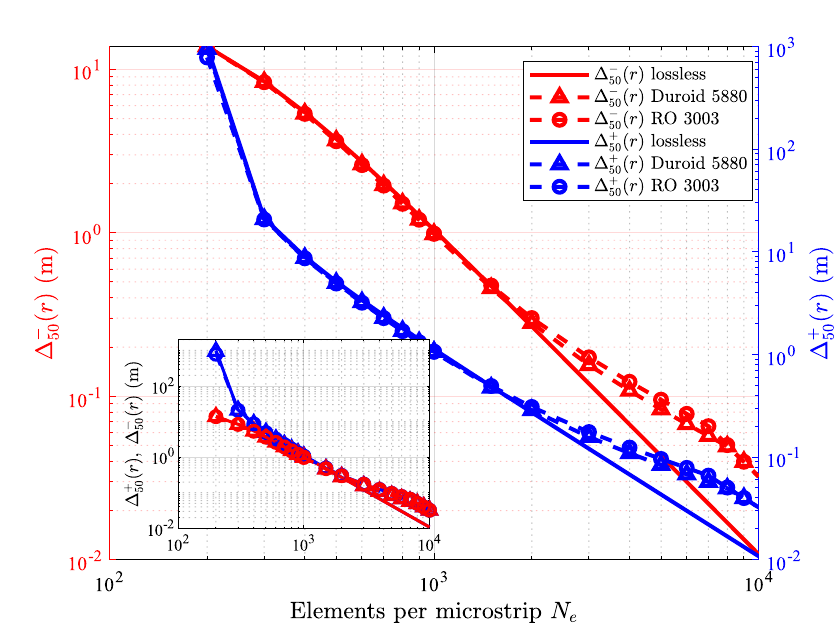}
    \caption{The range mismatch limits \(\Delta^{\pm}_{\kappa}(r)\) for \(\kappa=50\) are illustrated for three microstrip attenuation cases: i) lossless \(\alpha=0\) (solid lines), ii) Duroid 5880 with \(\alpha = 0.7381\,\text{m}^{-1}\) (dashed lines with triangle markers), and iii) RO 3003 with \(\alpha = 0.8629\,\text{m}^{-1}\) (dashed lines with circle markers). The comparison is conducted for increasing number of antenna elements per microstrip, \(N_e \in [200, 10^4]\), which effectively extends the DMA length, leading to more elements being significantly attenuated. The setup includes a DMA at the BS with \(N_m = 10\), \(d_e = d_m = \lambda/2\), where \(\lambda = 1\,\text{cm}\), and fixed values \(r = 28\,\text{m}\), \(\phi = \pi/4\) (rad), and \(z_0 = -0.5(N_e - 1)d_e = -0.5\,\text{m}\). The left vertical axis shows the values of \(\Delta^{-}_{\kappa}(r)\), while the right vertical axis shows \(\Delta^{+}_{\kappa}(r)\). For all considered cases, when \(N_e \geq 800\), the plus and minus range mismatch limits converge, that is, \(\Delta^{+}_{\kappa}(r) = \Delta^{-}_{\kappa}(r)\). This convergence is not clearly visible in the main plot due to the different scaling of the right and left vertical axes, but it becomes evident in the inset, where both limits are plotted against a common vertical axis.}
    \label{fig: Dr limits with attenuation}
\end{figure}

In Fig.~\ref{fig: Dr limits with attenuation}, we illustrate the \(\Delta^{\pm}_{\kappa}(r)\) limits with \(\kappa=50\%\), and \(N_e \in [200,10^4]\), under three scenarios: i) lossless (\(\alpha=0\)), ii) Duroid 5880 (\(\alpha=0.7381\)), and iii) RO 3003 (\(\alpha=0.8629\)). We use \(d_e=d_m=\lambda/2=0.5\) cm, \(r=28\) m, \(\phi=\pi/4\) rad, and \(z_0=-0.5(N_e-1) d_e = -0.5\) m. It is observed that for \(N_e \leq 2000\) (corresponding to a microstrip length of up to \(10\) m), the results for both lossless and lossy cases remain identical. However, beyond this point, attenuation effects become significant, causing the performance of lossy substrates to deviate from the ideal case and resulting in broader depth of focus limits. Among the considered materials, RO 3003 exhibits a slightly greater beam depth range compared to Duroid 5880 for the same number of elements, highlighting that stronger attenuation degrades near-field beamforming performance more rapidly.

\begin{figure}[t]
    \centering
    \includegraphics[width=1\textwidth]{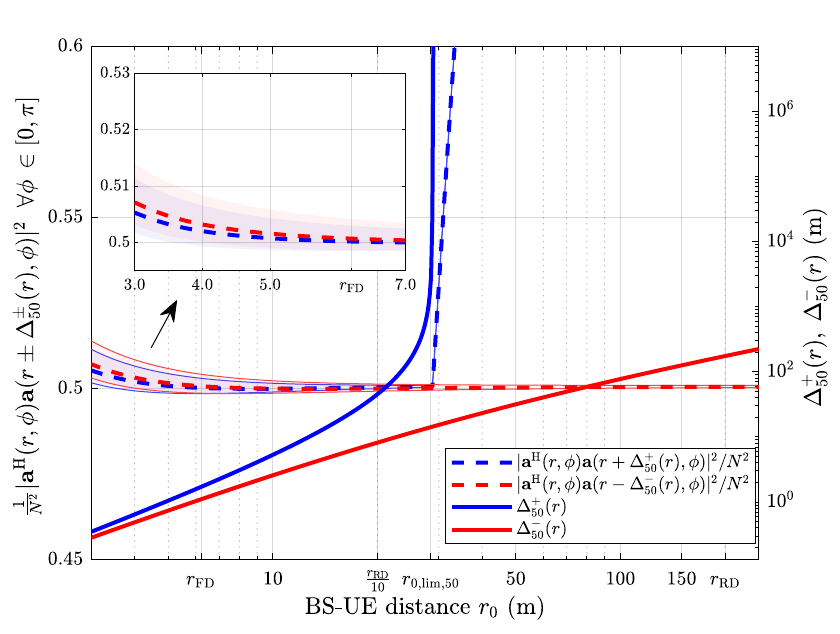}
    \caption{The relative beamforming gain for only range mismatch (left vertical axis) and the range mismatch limits \(\Delta^{\pm}_{\kappa}(r)\) in \eqref{eq: prop_Dr} for \(\kappa=50\) (right vertical axis) as functions of the BS-UE distance \(r_0\) in meters, considering a DMA at the BS with \(N_e=200\) and \(N_m = 10\) (i.e., \(N=2000\) elements), \(d_e=d_m=\lambda/2\) with \(\lambda = 1\,\text{cm}\), and \(z_0 = -0.5N_ed_e=-0.5\,\text{m}\). The blue and red error bars (i.e., shadowed areas) depict the range of values that the expressions \(\frac{1}{N^2} |\mathbf{a}^{\rm H}(r,\phi)\mathbf{a}(r \pm \Delta^{\pm}_{50}(r),\phi) |^{2}\) take \(\forall \phi \in [0,\pi]\), while the blue and red dashed lines depict the average values of these expressions with respect to \(\phi\). }
    \label{fig: Dr limits}
\end{figure}

Figure~\ref{fig: Dr limits} illustrates the \(\Delta^{\pm}_{\kappa}(r)\) range limits for the same DMA configuration as in Fig.~\ref{fig: Dr limits with attenuation}, but with a fixed \(N_e = 200\) and considering only the lossless scenario, which will be the focus of the remainder of this chapter. In this figure, the range limits are plotted with respect to increasing BS-UE distances \(r_0\), while the shadowed error bar shows where the values of these limits lie for all azimuth angles \(\phi \in [0,\pi]\). This is done to numerically validate the claim that, for DMAs with \( N_e \gg N_m \), the range limits are independent of the azimuth angle. This behavior is evident from the minor deviations observed, and for \( r_0 \leq 0.1 r_{\rm RD} \), the shadowed area converges to its mean value. Even in distances below the Fresnel limit \(r_0\in[3,r_{\rm FD}]\) that are shown in the inset figure, our analysis is still validated. It is noted, however, that, the distance from which our results hold depends on \(z_0\) (see \eqref{eq:Dist_approx}). The reader is referred to Fig.~3 in \cite{gavriilidis2025_beam_tracking} where \(z_0=1\) m; therein, the results become tight after \(r_{\rm FD}\). Furthermore, it can be observed from Fig.~\ref{fig: Dr limits} that, after the value \(r_{0,{\rm lim},50} \triangleq \sqrt{ r^2_{{\rm lim},50} + \left(z_0 + 0.5(N_e-1)d_e\right)^2 }\), \(\Delta^{+}_{\kappa}(r) \to \infty\) and the correlation factor exceeds \(0.5\), as expected since the UE has moved further away enough from the BS so that a limiting \(\Delta^{+}_{50}(r)\) distance does not exist. Finally, from the right vertical axis in Fig.~\ref{fig: Dr limits} it is shown that \(\Delta^{+}_{50}(r)\) has a steeper rate of change compared to \(\Delta^{-}_{50}(r)\), indicating that the depth of focus intervals become increasingly asymmetrical as \(r_{0}\) increases to the point where \(\Delta^{+}_{50}(r)\to \infty\) after \(r_{0}\geq r_{0,{\rm lim},50}\). This behavior is explained mathematically from the \(\Delta^{\pm}_{\kappa}(r)\) expression in \eqref{eq:Dr}, and its physical interpretation is that \(\Delta^{-}_{\kappa}(r)\) depicts the direction of UE movement towards the BS, where the near-field effects are stronger, hence, resulting in shorter depth of beam focus limits. 

In the following, we derive similar beamforming loss bounds considering a mismatch in the azimuth angle, i.e., \(\hat{\phi}=\phi \pm \Delta \phi\) \cite[Lemma~2]{gavriilidis2025_beam_tracking}.
\refstepcounter{graybox}
\label{box: Dphi beamfocusing}
\begin{svgraybox}
    {\textbf{Statement~\thegraybox:} When the BS beam focuses at the point \((r,\hat{\phi})\), the relative beamforming gain is derived as~\(\frac{G_{r,\hat{\phi}}}{G_{\rm opt}}\triangleq\frac{|\mathbf{a}^{\rm H}(r,\phi)\mathbf{a}(r  ,\phi \pm \Delta \phi) |^{2}}{N^2 } \approx \mathcal{L}^2(\zeta(\pm \Delta \phi))\), where function \(\mathcal{L}(x) \triangleq \frac{|\sin(x)|}{|N_m\sin(\frac{x}{N_m})|} \approx \frac{|\sin(x)|}{|x|} \) with \(\zeta(x) \triangleq N_m \frac{\pi d_m}{\lambda} \left(\cos(\phi)-\cos(\phi+x)\right)\).} 
\end{svgraybox}

Using the elaboration in Statement~\ref{box: Dphi beamfocusing}, we can derive the range of azimuth angles within which the beamforming gain is at least \(\kappa \%\) of its optimum value. To this end, we solve \(\mathcal{L}^2(\zeta(\pm \Delta \phi))=0.01{\kappa} \) with respect to $\zeta(\pm \Delta \phi)$ to obtain the solution \(\zeta_{\kappa}\). Function \(\mathcal{L}(\cdot)\) exhibits even symmetry and decays monotonically until reaching its first zero crossing \cite[Chap. 6.3]{balanis2016antenna}. Additionally, for small angular deviations \(\Delta \phi\), the approximation \(\zeta(\pm \Delta \phi) \approx \pm \zeta(\Delta \phi)\) holds, leading to the relation \(\mathcal{L}(\zeta(\pm \Delta \phi)) \approx \mathcal{L}(\zeta(\Delta \phi))\). Among possible solutions, we retain the smallest positive root, denoted by \(\zeta_{\kappa}\). We then assign \(\zeta(\pm \Delta \phi) = \zeta_{\kappa}\), yielding the subsequent expression
\eqref{eq:Dphi}:
\begin{equation}\label{eq:Dphi}
    \Delta_{\kappa} (\phi) = \left| \frac{\zeta_{\kappa}\lambda }{\pi  N_m d_m  \sin(\phi)}\right|.
\end{equation}
The latter solution can be now used to derive the angle width for which it holds that \( \frac{G_{r,\hat{\phi}}}{G_{\rm opt}} =\frac{|\mathbf{a}^{\rm H}(r,\phi)\mathbf{a}(r  ,\hat{\phi}) |^{2}}{N^2 } \geq \kappa \%\) for \(\hat{\phi} \in [\phi - \Delta_{\kappa} (\phi) , \phi + \Delta_{\kappa} (\phi)]\). It is noted that, to enhance accuracy and circumvent singularities at \(\phi = \{0,\pi\}\), one can directly solve the equation \(|\cos(\phi) - \cos(\phi \pm \Delta \phi)| = \frac{\zeta_k \lambda}{\pi N_m d_m}\) numerically, omitting the use of the Taylor series approximation that is implied in \eqref{eq:Dphi}.
It can be deduced from \eqref{eq:Dphi} that this angle width becomes larger as  \(\phi \to 0\), i.e., when the UE lies to the side of the DMA, and narrower as \(\phi \to \pi/2\). The latter case indicates that the UE is located at the broadside of the DMA. This follows intuitively from the fact that an antenna array typically achieves maximum directivity in its broadside direction, while the lobes created to its sides are wider.

We next present a closed-form approximation for the relative beamforming gain for simultaneous mismatches in the range \(r\) and the azimuth angle \(\phi\) \cite[Lemma~3]{gavriilidis2025_beam_tracking}.

\refstepcounter{graybox}
\label{box: Dr and Dphi beamfocusing}
\begin{svgraybox}
    \textbf{Statement~\thegraybox:} For BS beam focusing at the point \((\hat{r},\hat{\phi})\), the relative beamforming gain is~\(\frac{G_{\hat{r},\hat{\phi}}}{G_{\rm opt}}\triangleq\frac{|\mathbf{a}^{\rm H}(r,\phi)\mathbf{a}(r \pm \Delta r ,\phi \pm \Delta \phi) |^{2}}{N^2 } \approx \mathcal{M}^2(a(\pm \Delta r),\zeta(\pm \Delta \phi)) ,\) with \(\mathcal{M}(x,y) \triangleq I(x) \mathcal{L}(y)  \).
\end{svgraybox}

\begin{figure}[t]
    \centering
    \includegraphics[width=1\textwidth]{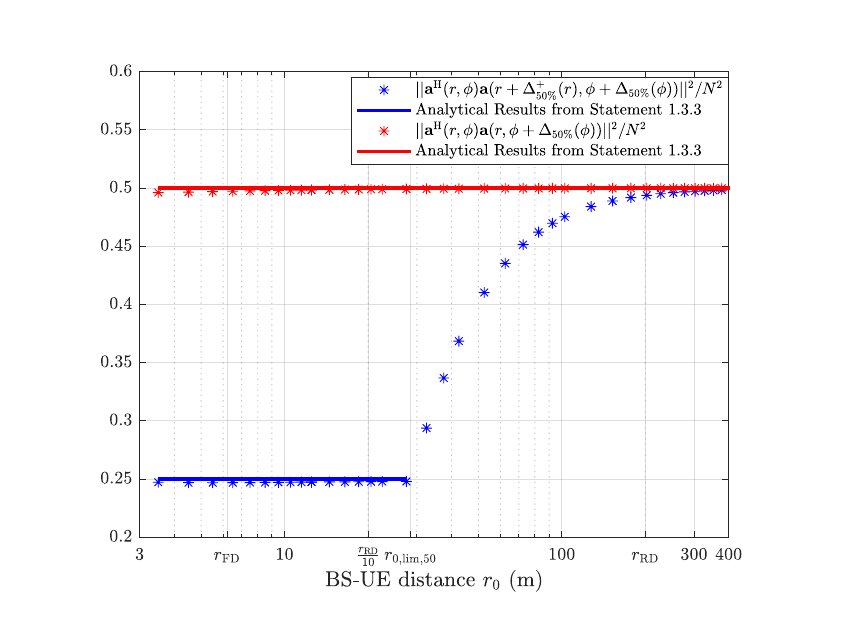}
    \caption{The relative beamforming gains for only azimuth angle mismatch (Statement~\ref{box: Dphi beamfocusing}) and for both range and angle mismatches (Statement~\ref{box: Dr and Dphi beamfocusing}) versus the BS-UE distance $r_0$ (m) for $\phi=\pi/4$ (rad) and using the same DMA parameters as in Fig.~\ref{fig: Dr limits}.}
    \label{fig: Dr and Dphi limits}
\end{figure}

The analytical expressions for the relative beamforming gain corresponding to the case studies in Statement~\ref{box: Dphi beamfocusing} (azimuth mismatch \(\Delta_{\kappa} (\phi)\)) and Statement~\ref{box: Dr and Dphi beamfocusing} (combined range and azimuth mismatches \(\Delta^{\pm}_{\kappa}(r)\) and \(\Delta_{\kappa} (\phi)\)) are validated in Fig.~\ref{fig: Dr and Dphi limits} for \(\kappa = 50\), using the same DMA configuration at the BS as in Fig.~\ref{fig: Dr limits}. The results confirm that the analytical predictions are consistent with the observed performance: when a \(\Delta_{50}(\phi)\) mismatch occurs in the azimuth coordinate, the beamforming gain drops to half of the optimal value (\(0.5\) in Fig.~\ref{fig: Dr and Dphi limits}); when both \(\Delta^{+}_{50}(r)\) and \(\Delta_{50}(\phi)\) mismatches occur, the gain reduces to one quarter of the optimal, losing half due to each mismatch (\(0.25\) in Fig.~\ref{fig: Dr and Dphi limits}). Moreover, it is again shown that once the range limit \(r_{0,\lim,50}\) is exceeded, the degradation due to the range mismatch begins to subside, and beyond \(r_{\rm RD}\), only the azimuth mismatch contributes to gain deterioration. This decoupled analytical characterization in both the range \(r\) and azimuth angle \(\phi\) of the beamforming gain achieved by a DMA-based transmitter will be instrumental in the sequel, where we derive the worst-case direction of UE movement with respect to near-field beamforming performance.

\subsection{Effective Beam Coherence Time}\label{subsec: Beam Coherence Time}
We next study the degradation of beamforming gain as the UE moves across the 2D plane from a given starting position. Specifically, for a specified displacement in meters, we determine the trajectory that results in the steepest drop in beamforming gain, by using the analytical function from Statement~\ref{box: Dr and Dphi beamfocusing}. As we will demonstrate, this analysis enables the derivation of the minimum duration required for the UE to experience a particular loss relative to the theoretical maximum \(G_{\rm opt}\). We denote this interval as the \textit{effective beam coherence time}, which will later serve as the foundation for defining dynamic estimation intervals, based on the duration for which the gain remains above a designated threshold.

\subsubsection{Direction of Maximum Beamforming Gain Deterioration}\label{sec:direction_of_fastest_BF_deterioration}
Assume that the UE travels a distance \(c\) from its previous location \(\mathbf{p}_{t-1}=[r_{t-1},\phi_{t-1}]\) at time \(t-1\), moving within the \(xy\)-plane. From the cosine law, its new position can be expressed as \(\mathbf{p}_t=\left[r_{t-1}+d,\, \phi_{t-1} \pm \arccos\left(1-\frac{c^2-d^2}{2r^2+2rd}\right)\right]\) for \(|d|\leq c\), meaning the trajectory lies along a circle of radius \(c\) centered at \(\mathbf{p}_{t-1}\). At the extreme case \(d=0\), the angular mismatch reaches a maximum value \(|\mathcal{D} \phi_{\max}|=\arccos\left(\frac{2r^2-c^2}{2r^2}\right)\), whereas at \(d = \pm c\), the largest radial variation is observed, i.e., \(|\mathcal{D} r_{\max}|=c\).

By expressing the mismatch functions \(a(x)\) and \(\zeta(y)\), defined in Statements~\ref{box: Dr beamfocusing} and~\ref{box: Dphi beamfocusing}, as functions of \(d\), with \(x = d\) and \(y(d) \triangleq \arccos\left(1-\frac{c^2-d^2}{2r^2+2rd}\right)\), we reformulate the gain function \(\mathcal{M}(\cdot,\cdot)\) from Statement~\ref{box: Dr and Dphi beamfocusing} as a function of \(d\), for a given distance \(c\), as follows:
\begin{equation}\label{eq:BF_gain_for_movement_d}
\mathcal{M}(d) =
\begin{cases}
    I(a(d))\,\mathcal{L}(\zeta(y(d))), &  d \in (0,c] \\
    I(a(d))\,\mathcal{L}(\zeta(y(d))), & d \in [-c,0]
\end{cases}.
\end{equation}
This two-branch representation highlights the discontinuity of \(\mathcal{M}(d)\) at \(d=0\), introduced by the absolute value inside \(a(d)\). To locate the direction that causes the most severe beamforming gain loss, we identify the critical points of \(\mathcal{M}(d)\) over the interval \([-c,0]\): \textit{i}) the roots of the first derivative \(\frac{{\rm d}\mathcal{M}(d)}{{\rm d}d}=0\); and \textit{ii}) the interval boundaries. Only the second branch is studied due to the fact that \(a(-|d|)>a(|d|)\Rightarrow I(a(-|d|))<I(a(|d|))\) as well as \(y(-|d|)>y(|d|)\Rightarrow\zeta(y(-|d|))>\zeta(y(|d|)) \Rightarrow \mathcal{L}(\zeta(y(-|d|)))<\mathcal{L}(\zeta(y(-|d|)))\), indicating that the minimum exists in the second branch (\(\rm acos(\cdot)\) and \(\zeta(\cdot)\) are decreasing and increasing functions, respectively). By evaluating \(\mathcal{M}(d)\) at these candidate points, we obtain \(d_{\min} \triangleq \arg\min_{|d|\leq c} \mathcal{M}(d)\), and the associated position is
\[
\mathbf{p}_{t,\min}=\left[r_{t-1}+d_{\min},\, \phi_{t-1} \pm \arccos\left(1-\frac{c^2 - d_{\min}^2}{2r^2 + 2rd_{\min}}\right)\right],
\]
which corresponds to the direction that causes the most rapid gain degradation, i.e., from \(\mathbf{p}_{t-1}\) to \(\mathbf{p}_{t,\min}\). Our next objective is to quantify the minimum distance \(c_{\min,\kappa\%}\) that the UE must move for the beamforming gain to degrade to \(\kappa\%\) of its optimal value \(G_{\rm opt}\). To avoid computationally intensive operations such as numerical root-finding or iterative line searches, we adopt the approximation
\( c_{\min,\kappa \%} \approx \min \left\{\Delta^{-}_{\kappa}(r),\, 2r\sin\left(0.5\Delta_{\kappa} (\phi)\right)\right\},\)
which has been shown in~\cite{gavriilidis2025_beam_tracking} to closely approximate the exact value. Letting \(u\) denote the UE's speed, we define the \textit{effective beam coherence time} as \(T_{{\rm c},\kappa \%} \triangleq \frac{c_{\min,\kappa\%}}{u}\), which represents the minimum time interval over which the UE can move before the beamforming gain falls below \(\kappa\%\) of \(G_{\rm opt}\). This metric will be subsequently employed in the design of our beam tracking protocol.

\section{Sensing-aided Near-Field Beam Tracking}\label{sec:Proposed_BA}
In this section, we present the proposed sensing-aided near-field beam tracking framework, which builds upon the analytical findings established thus far. A central component of this framework is the design of a polar, dynamic, and non-uniform coordinate grid that enables efficient UE localization by minimizing the required number of samples, while accounting for the spatially varying near-field effects within the search region.

\subsection{Localization Design Objective}
Given an estimate of the UE’s position at time \(t-1\), denoted by \(\mathbf{\hat{p}}_{t-1} = [\hat{r}_{t-1}, \hat{\phi}_{t-1}]\), along with the elapsed time interval \(T\) and a coarse estimate of the UE's velocity magnitude, we estimate the displacement \(\hat{c}\) traversed by the UE. Accordingly, the search region for the updated position \(\mathbf{\hat{p}}_t = [\hat{r}_t, \hat{\phi}_t]\) is defined as the circular area \({\rm C}(\mathbf{\hat{p}}_{t-1}, \hat{c})\) centered at \(\mathbf{\hat{p}}_{t-1}\) with radius \(\hat{c}\), lying on the UE’s motion plane. The localization objective is then formulated as maximizing the beamforming gain, i.e., identifying the pair \((\hat{r}, \hat{\phi})\) such that the gain \(G_{\hat{r},\hat{\phi}}\) approaches the optimal value \(G_{\rm opt}\). This can be modelled as the following optimization problem:
\begin{align*}
    \mathcal{OP}_2: \quad \max_{\hat{r}, \hat{\phi}} \,\, |\mathbf{a}^{\rm H}(\hat{r}, \hat{\phi}) \mathbf{h}_{\rm LoS}| \quad \text{s.t.} \quad (\hat{r}, \hat{\phi}) \in {\rm C}(\mathbf{\hat{p}}_{t-1}, \hat{c}).
\end{align*}

Due to the use of a DMA-based receiver at the BS, problem \(\mathcal{OP}_2\) cannot be directly solved based on the received pilot signal \(\mathbf{y}_b\) in \eqref{eq:DMA_UE}. To address this, we propose receiving the UE's pilot symbols through a set of different beam focusing matrices \(\mathbf{\bar{Q}}\), which will be detailed later. This strategy is inspired by the Simultaneous Orthogonal Matching Pursuit (SOMP) approach described in \cite[Alg.~3]{CS_DoA}, relying on the identification of a dominant spatial component (i.e., knowledge of only the strongest path is required). We also assume that the beamforming gain \(|\mathbf{a}^{\rm H}(\hat{r}, \hat{\phi}) \mathbf{h}|\) is maximized when the coordinates correspond to the LoS path. This assumption holds because beam sweeping is done to a specific area of interest in each estimation step, and scatterers located outside this region do not impact the localization. For scatterers within this area, that is, within \({\rm C}(\mathbf{\hat{p}}_{t-1}, \hat{c})\), we assume that their power is always lower than that of the LoS component due to higher path loss. Consequently, solving \(\max_{\hat{r},\hat{\phi}} \; |\mathbf{a}^{\rm H}(\hat{r},\hat{\phi}) \mathbf{h}|\) provides a valid estimate of the UE’s coordinates.

\subsection{Dynamic Non-Uniform Coordinate Grid}\label{subsec:Dynamic_sampling}
To efficiently solve \(\mathcal{OP}_2\), we now focus on constructing the minimum necessary set of sampling points needed to estimate the UE coordinates \((\hat{r},\hat{\phi})\) that yield a satisfactory fraction of the optimal beamforming gain \(G_{\rm opt}\). This design leverages the beamforming analysis developed in Section~\ref{suubsec: BF_gain_mmismatch}. Specifically, we introduce a dynamic grid tailored to the time-varying region of interest \({\rm C}\left(\mathbf{\hat{p}}_{t-1},\hat{c}\right)\), which encapsulates all possible UE positions at time \(t\).

The sampling resolution used in the search for the UE position depends on the desired proximity to the optimal beamforming gain. To that end, we propose a dynamic, non-uniform sampling scheme that accepts as input the region \({\rm C}\left(\mathbf{\hat{p}}_{t-1},\hat{c}\right)\) and a target gain threshold \(\delta \%\) of \(G_{\rm opt}\), and returns a set of sampling points sufficient to ensure that this target is met. The sampling procedure is described in Alg.~\ref{alg:sampling_proc} and operates as follows. The key idea is that each sampling point \((r_i,\phi_j)\) defines a decision region given by \([r_i - \Delta^{-}_{\delta}(r_i), r_{i}+ \Delta_{\delta}^{+}(r_i)] \cup [\phi_j - \Delta_{\delta}(\phi_j), \phi_j + \Delta_{\delta}(\phi_j)]\), within which it holds that for all \((\hat{r},\hat{\phi})\), the normalized gains \(\frac{G_{\hat{r},\phi_j}}{G_{\rm opt}}\) and \(\frac{G_{r_i,\hat{\phi}}}{G_{\rm opt}}\) are at least \(\delta \%\). The decision regions are non-overlapping and together span the entire area \({\rm C}\left(\mathbf{\hat{p}}_{t-1},\hat{c}\right)\).

\begin{figure}[t]
    \centering
    \includegraphics[width=1\textwidth]{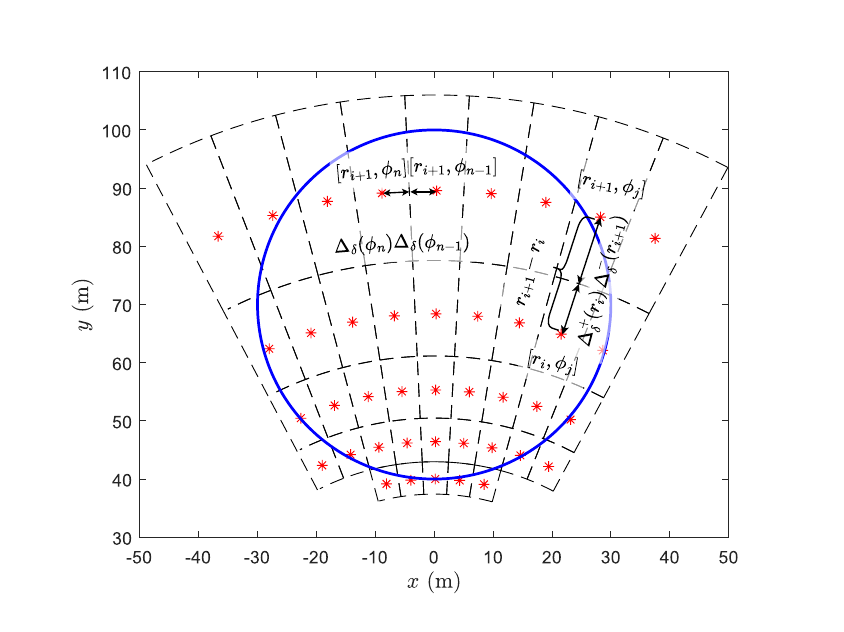}
    \caption{The proposed non-uniform coordinate grid on the $xy$ plane where the UE moves, as implemented via Algorithm~\ref{alg:sampling_proc}, for a central point at the polar coordinates \(\hat{r}=70\) ${\rm m}$ and \(\hat{\phi} =\pi/2\), resolution \(\delta\%=80\%\), and radius \(\hat{c}=30\) ${\rm m}$. It is noted that, since \(\sin(\phi_j) \to 0\) when \(\phi_j \to 0\), the angular distances become wider in that regime, following the derived angular limits in \eqref{eq:Dphi}, while for the presented radial distances in \eqref{eq:Dr}, \(\Delta^{\pm}_{\delta}(r_i)\) increases with increasing $r$. The parameters of the DMA-based BS considered are the same as in Figs.~\ref{fig: Dr limits}-\ref{fig: Dr and Dphi limits}.}
    \label{fig:sampling}
\end{figure}

To construct such a grid, Algorithm~\ref{alg:sampling_proc} samples the region \({\rm C}\left(\mathbf{\hat{p}}_{t-1},\hat{c}\right)\) with resolution \(\delta \%\), such that the spacing between two consecutive radial samples \(r_i\) and \(r_{i+1}\), with \(r_{i+1} \geq r_i\), is approximately \(2\Delta^{+}_{\delta}(r_i)\) (Steps~12 and 13). Similarly, the angular spacing between consecutive azimuth samples \(\phi_j\) and \(\phi_{j+1}\) is set to approximately \(2\Delta_{\delta}(\phi_j)\) (Steps~5 and 6). Step~15 further ensures that for each radial level \(r_{i}\), only azimuth angles within the region \({\rm C}(\mathbf{\hat{p}}_{t-1},\hat{c})\) are considered. As an illustration, in the sampling example of Fig.~\ref{fig:sampling}, when constructing sampling points at the smallest range \(r_1 = r - \hat{c}\) (Step~9), only \(5\) azimuth angles are selected (first arc), while at the next radial level, \(9\) are chosen. Collectively, the union of the decision regions of all selected points fully covers \({\rm C}\left(\mathbf{\hat{p}}_{t-1},\hat{c}\right)\).

Consequently, the error of the on-grid estimation is bounded such that the gain loss is always less than \(1 - \delta\%\) in both the radial and angular dimensions. For any UE location \((r_t,\phi_t) \in {\rm C}(\mathbf{\hat{p}}_{t-1},\hat{c})\), there exists a unique sampling point, denoted \((\hat{r}_t,\hat{\phi}_t)\), that serves as the next position estimate and satisfies
\[
\frac{G_{\hat{r}_{t},\phi_t}}{G_{\rm opt}},\quad \frac{G_{r_t,\hat{\phi}_{t}}}{G_{\rm opt}} \geq \delta \%.
\]
In the worst-case scenario, however, that is when the UE lies exactly at the edge of a decision region with \(r_t = \hat{r}_t \pm \Delta_{\delta}^{\pm}(\hat{r}_t)\) and \(\phi_t = \hat{\phi}_t \pm \Delta_{\delta}(\hat{\phi}_t)\), the corresponding gain becomes:
\[
\mathcal{M}^2\left(a\left(\pm \Delta_{\delta}^{\pm}(\hat{r}_t)\right), \zeta\left(\pm \Delta_{\delta}(\hat{\phi}_t)\right)\right) = (\delta \%)^2,
\]
as suggested in Statement~\ref{box: Dr and Dphi beamfocusing}. Therefore, to guarantee a minimum gain of \(\delta \%\) in this edge case, one would need to sample the region with resolution \(\sqrt{\delta \%}\). However, such an approach is inefficient, as it significantly increases the sampling density to satisfy a rare boundary case. In practice, sampling with \(\delta \%\) resolution already ensures an average gain greater than \(\delta \%\), as shown in \cite[Lemma~4]{gavriilidis2025_beam_tracking}. In conclusion, Algorithm~\ref{alg:sampling_proc} dynamically adapts the sampling resolution based on the current UE position estimate \(\mathbf{\hat{p}}_{t-1}\) and the anticipated displacement \(\hat{c}\), thereby ensuring that a gain of at least \(\delta\%\)\(G_{\rm opt}\) is achieved.

\begin{algorithm}[!t]
\caption{\textcolor{black}{Dynamic Non-Uniform Coordinate Grid Design}}\label{alg:sampling_proc}
 \textbf{Input:} \((\hat{r}_{t-1},\hat{\phi}_{t-1})\), \(\hat{c}\), and desired localization resolution \(\delta \%\).\\ \vspace{-3 mm}
\begin{algorithmic}[1]
\State Compute \(\mathcal{D} \phi_{\max}={\rm acos}\left(\frac{2\hat{r}^2_{t-1}-\hat{c}^2}{2\hat{r}^2_{t-1}}\right)\).
\State Initialize \( s_{\phi} = 1\) and \(\phi_{s_{\phi}} = \hat{\phi}_{t-1}-\mathcal{D} \phi_{\max}\).
\While{\(\phi_{s_\phi}+\Delta_\delta(\phi_{s_\phi}) \leq \hat{\phi}_{t-1}+ \mathcal{D} \phi_{\max}\)}
\State  \(s_\phi = s_\phi +1\).
\State  \(\phi_{\rm temp}= \phi_{s_{\phi}-1} + \Delta_\delta(\phi_{s_{\phi}-1})\).
\State  \( \phi_{s_\phi} =\phi_{\rm temp} + \Delta_\delta(\phi_{{\rm temp}})\).
\EndWhile
\State Set $S_\phi=s_\phi$ and define \(\bm{\phi}=[\phi_{1},\phi_2,\dots,\phi_{S_\phi}]\).
\State Initialize \(s_r = 1\) and \(r_{s_r} = \hat{r}_{t-1} - \hat{c}\).
\While{\(r_{s_r} + \Delta_{\delta}^{+}\!\left(r_{s_r}\right) \leq \hat{r}_{t-1} + \hat{c}\)}
\State \(s_{r} = s_r+1\).
\State \(r_{\rm temp} = r_{s_r-1} + \Delta^{+}_{\delta}\! (r_{s_r-1})\).
\State \(r_{s_r} = r_{\rm temp} + \Delta^{+}_{\delta}\!\left(r_{\rm temp}\right) \).
\State Compute \(\mathcal{D} \phi_{\max, s_r} =  {\rm acos}\left( \frac{r^2_{s_r}+ \hat{r}^2_{t-1}-\hat{c}^2}{2 r_{s_r} \hat{r}_{t-1}} \right)\).
\State Formulate \(\boldsymbol{\phi}_{s_r}\!=\!\big\{\boldsymbol{\phi}\!\mid\! [\phi_{i}\!+\!\Delta_{\delta}(\phi_i) \!\geq\!\hat{\phi}_{t-1}\!-\!\mathcal{D}\phi_{\max, s_r}] \;\!\cap\)

\hspace{-4mm}
\(\; [\phi_{i}\!-\!\Delta_{\delta}( \phi_i) \!\leq\!  \hat{\phi}_{t-1}\!+\!\mathcal{D} \phi_{\max, s_r}],\,i=1,2,\ldots,S_\phi\big \}\).
\State Compute \(S_{\boldsymbol{\phi}_{s_r}} = {\rm length}\left(\boldsymbol{\phi}_{s_r}\right)\).
\EndWhile
\State Set $S_r=s_r$.
\end{algorithmic}
\textbf{Output:} Coordinate grid \((r_{s_r},[\boldsymbol{\phi}_{s_r}]_i)\) \(\forall s_r\! =\! 1,2,\ldots, S_r\) and \(\forall i\! =\! 1,2,\ldots,S_{{\boldsymbol{\phi}}_{s_r}}\) for time $t$.
\end{algorithm}

\subsubsection{Hybrid Beam Sweeping}\label{subsec:Hybrid_Combining}
As shown in Fig.~\ref{fig:sampling}, to sample the UE’s circular area of interest at each estimation phase, we apply Algorithm~\ref{alg:sampling_proc} to generate a grid composed of arcs with constant range \(r\). Within each arc, we then search over the corresponding azimuth grid points to estimate the angle \(\phi\). From the approximation in~\eqref{eq:Dist_approx}, it becomes evident that this gridding and search process can be implemented in an over-the-air fashion by configuring the analog combiner to focus on a specific radial distance \(r\), while performing digital scanning over azimuth angles \(\phi\) to identify the direction that maximizes the received signal power.

To this end, we define the hybrid beamforming configuration through an analog combiner \(\mathbf{\bar{Q}}_{r}\) and a digital combiner \(\mathbf{v}_{\phi}\), which jointly focus on a spatial direction \((r,\phi)\). These are defined for all \(i=0,\ldots,N_m-1\) and \(n=0,\ldots,N_e-1\) as:
\begin{align}
    & [\mathbf{{\bar{Q}}}_{r}]_{iN_e+n,i} \!=\! \frac{\jmath \! + \! e^{-\jmath \left(\!k \!\left(\! \frac{i_x^2d_m^2+n^2 d_e^2}{2r} + \frac{z_0 n d_e}{r} \right) {\rm -}\beta \rho_{i,n}\right)}}{2}, \label{eq:Q_r}\\
    & [\mathbf{v}_{\phi}]_i = e^{\jmath k \left(i_x d_m\cos(\phi) + \frac{i_x^2 d_m^2}{2r}\cos^2(\phi)\right)}.\label{eq:v_phi}
\end{align}
Neglecting the microstrip line propagation effects within the DMA, the combined response simplifies to\footnote{In this case, since we are operating at the receiver side, no normalization factor is applied, unlike the transmit case in~\eqref{eq: Hybrid Analog and Digital Config for Los focusing}, where power constraints must be satisfied.} \(\mathbf{v}_{\phi}^{\rm H}\mathbf{\bar{Q}}^{\rm H}_r = 0.5\,\mathbf{a}^{\rm H}(r,\phi)\). This hybrid beam scanning process is integrated into our localization framework as follows. At each estimation phase, we configure the analog sensing matrix \(\mathbf{\bar{Q}}_{r}\) to sequentially focus on the radial samples \(r\) defined by the non-uniform polar grid. This results in the received signal:
\begin{equation}
    \mathbf{y}_{b,r}\triangleq\mathbf{\bar{Q}}_{r}^{\rm H}\mathbf{h}x_u + \mathbf{\bar{Q}}_{r}^{\rm H} \mathbf{n}_b,
\end{equation}
as per~\eqref{eq:DMA_UE}. Then, for each fixed \(r\), we digitally scan across the angular grid points \(\phi\) and select the one maximizing the matched filter output \(|\mathbf{v}^{\rm H}_\phi\mathbf{y}_{b,r}|\). This procedure effectively solves the localization problem \(\mathcal{OP}_2\) under the common assumption in channel estimation studies, that the UE remains stationary throughout the beam search interval~\cite{CS_DoA,CS_NF_RIS}.

When the search radius \(\hat{c}\) in Algorithm~\ref{alg:sampling_proc} becomes large, the estimation overhead can be mitigated by employing CS techniques. These methods construct a dictionary composed of focusing vectors that are mutually orthogonal in both angular and radial domains~\cite{Appendix_approx,CS_NF_RIS,Position_Est_mmWave}. A practical approach toward that end could be initially executing Algorithm~\ref{alg:sampling_proc} at a coarse resolution, specified by a small \(\bar{\delta}\% \ll 1\), to determine the coordinate pair \([\bar{r}, \bar{\phi}]\) corresponding to the highest correlation response. Subsequently, a finer resolution search with step size \(\delta'\%\) can be conducted within a localized region\footnote{This refined sampling step may also be performed without confining the search to a circular area by skipping Steps $14$ and $15$ of Algorithm~\ref{alg:sampling_proc}.} defined by \(r \in [\bar{r} - \Delta^{-}_{\bar{\delta}}(\bar{r}), \bar{r} + \Delta^{+}_{\bar{\delta}}(\bar{r})]\) and \(\phi \in [\bar{\phi} - \Delta_{\bar{\delta}}(\bar{\phi}), \bar{\phi} + \Delta_{\bar{\delta}}(\bar{\phi})]\). The latter part of the procedure provides an off-grid estimation, which can also be carried out via higher complexity algorithms (e.g., via likelihood function maximization)~\cite{DMA_loc_Nir,RIS_localisation_George_henk_ML_estimator,Appendix_approx}.

\subsection{Near-Field Beam Tracking Protocol and Algorithm}\label{subsec:beam_tracking}

\begin{figure}[t]
    \centering
    \includegraphics[width=1\textwidth]{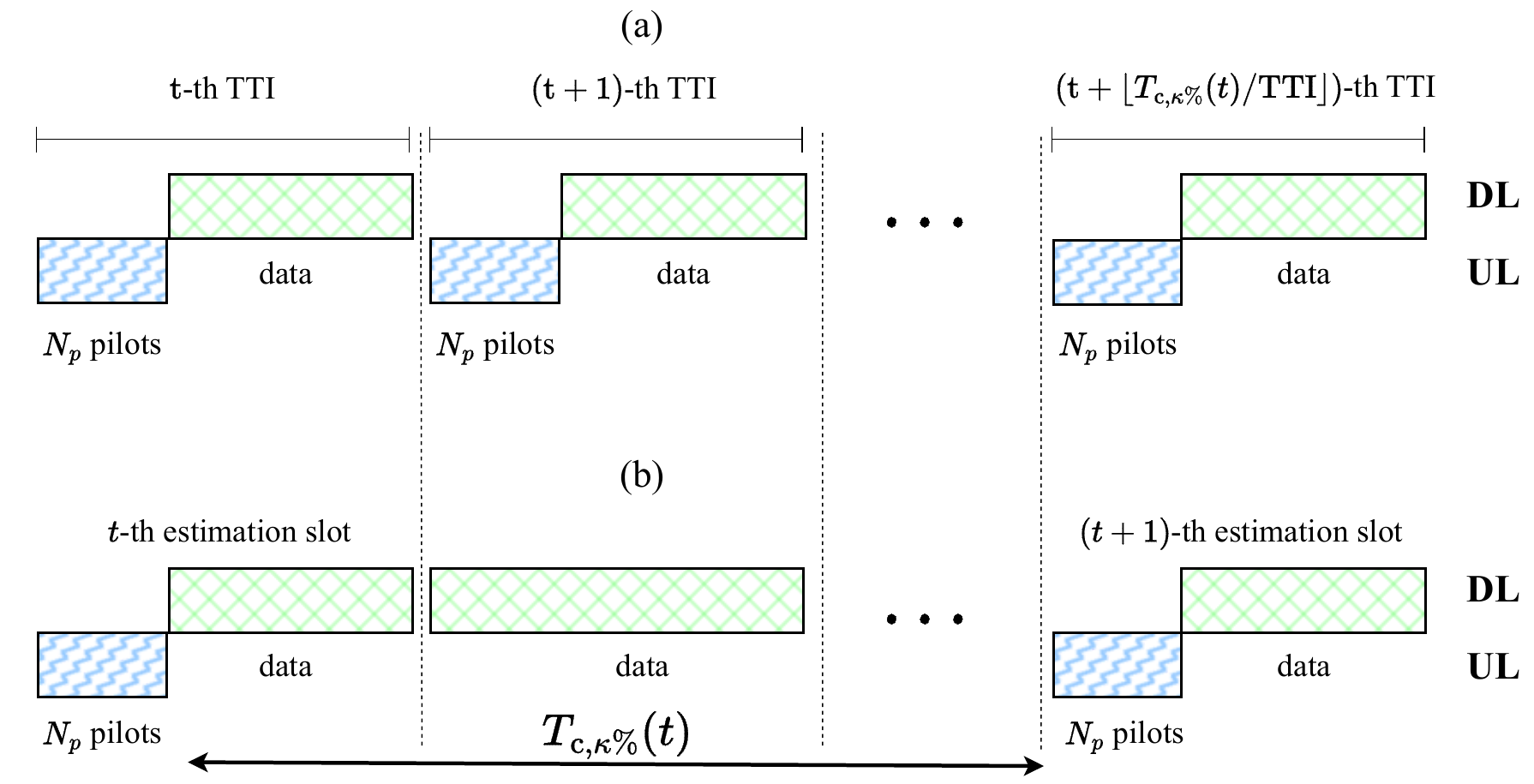}
    \caption{UL and DL in TDD: (a) Conventional communication protocol where channel estimation is conducted at every TTI; and (b) Proposed communication protocol where channel estimation is triggered based on the effective beam coherence time, which varies dynamically with respect to the beamforming loss parameter \(\kappa\) and the UE's state (position and velocity). In this illustration, \(T_{{\rm c},\kappa\%}(t)\) denotes the time interval between the \(t\)-th and \((t+1)\)-th estimation instances under the proposed near-field beam tracking scheme, with the first occurring during the \(t\)-th TTI and the second during the \((t{+}\lfloor{T_{{\rm c},\kappa\%}(t)/{\rm TTI}}\rfloor)\)-th TTI. If an outage occurs (e.g., due to a blocked LoS path), the UE initiates a channel estimation request before the expiration of \(T_{{\rm c},\kappa\%}(t)\).}
    \label{fig:UL_protocol}
    \vspace{-2mm}
\end{figure}

When the BS performs a UE position estimation to design its hybrid beam focusing configuration, it can also compute the minimum time required for the DL beamforming gain to drop to \(\kappa\%\) of its optimal value. This was previously defined as the effective beam coherence time \(T_{{\rm c}, \kappa \%}\). We propose to exploit this information at the BS to request UL pilot transmissions from the UE, thereby proactively initiating near-field beam sweeping. In this way, beam sweeping occurs every \(T_{{\rm c}, \kappa \%}\), rather than at every Transmission Time Interval (TTI)  or upon outage detection from the UE side, as is typical in conventional Time Division Duplexing (TDD) systems. The parameter \(\kappa\) allows the protocol to adapt its update interval based on the required Quality of Service (QoS) of the link. The proposed near-field beam tracking protocol is illustrated in Fig.~\ref{fig:UL_protocol}.

As will become evident in the sequel, the value of \(T_{{\rm c}, \kappa \%}\) can vary between successive estimation intervals. Notably, dense pilot signaling is employed when the UE is located close to the BS, where the beamforming gain is highly sensitive to position errors. Conversely, when the UE is farther from the BS, only coarse position estimates are required, as deviations from the focusing point have a limited effect on the link’s QoS. This behavior stems from the dependency of \(\mathcal{M}(\cdot)\) in~\eqref{eq:BF_gain_for_movement_d} on \(r\) for a movement of \(c\) meters. Specifically, for \(r \gg c\), we have \(a(d) \to 0\), which yields \(I(a(d)) \to 1\), \(y(d) \to 0\) for all \(d\), and therefore \(\mathcal{L}(\zeta(y(d))) \to 1\), implying \(\mathcal{M}(d) \to 1\).

To compute \(T_{{\rm c}, \kappa \%}\), knowledge of the UE velocity magnitude is required. However, this value is typically unknown and not constant over time. One way to estimate this quantity \(\hat{u}_t\) at each estimation slot \(t\) (see Fig.~\ref{fig:UL_protocol} for the relation between estimation slots and TTIs) is through a geometrically weighted moving average with \(\gamma > 1\) and \(w_j \triangleq \frac{\gamma^j}{\gamma^t - 1}\):
\begin{equation}\label{eq:predict_velocity}
    \hat{u}_t = \max\left\{ \sum_{j=0}^{t-1} w_j \hat{u}_j,\, u_{\rm th} \right\},
\end{equation}
where \(\hat{u}_{j+1} = \|\mathbf{\hat{p}}_{j+1} - \mathbf{\hat{p}}_j\|_2 / T_{{\rm c}, \kappa \%}(j)\), and \(T_{{\rm c}, \kappa \%}(j)\) is the effective beam coherence time computed at the end of the \(j\)-th estimation slot. The threshold \(u_{\rm th}\) ensures that \(\hat{u}_t\) remains non-zero.

Algorithm~\ref{alg:tracking} summarizes the near-field beam tracking strategy at each estimation slot \(t\). The BS computes the UE position estimate \(\mathbf{\hat{p}}_t\), the velocity magnitude estimate \(\hat{u}_t\), and the effective beam coherence time \(T_{{\rm c}, \kappa \%}(t)\), which will determine when the next estimation (at slot \(t+1\)) is performed. The BS then focuses its beam toward \(\mathbf{\hat{p}}_t\) (as described in \eqref{eq: Hybrid Analog and Digital Config for Los focusing}), and maintains this configuration until \(T_{{\rm c}, \kappa \%}(t)\) elapses and the next estimation is triggered. To account for errors in the prediction and estimation of the UE movement, we introduce robustness factors \(e_c\) and \(e_u\), which increase the expected displacement and velocity from \(c\) to \(c(1 + e_c)\) and from \(\hat{u}_t\) to \(\hat{u}_t(1 + e_u)\), respectively. This ensures more conservative updates under uncertainty. To obtain the initial estimates in the absence of prior information, conventional high-complexity estimation schemes can be employed, such as~\cite{Wideband_Hybrid_Tracking}.

Regarding the pilot length required, we note that, in Algorithm~\ref{alg:tracking}, the total number of analog combiner reconfigurations is equal to the number of radial samples \(S_r\). Furthermore, the UE area of interest lies within a radius of \(c_{\kappa \%}\), and as \(r_0\) increases, \(c_{\kappa \%} \leq \Delta^{-}_{\kappa}(r)\) increases accordingly. The following remark establishes that, within the presented beam tracking framework—where the search area is defined via \(c_{\kappa \%}\) and sampling is carried out using the non-uniform coordinate grid of Algorithm~\ref{alg:sampling_proc}—an upper bound on \(S_r\) exists.

\refstepcounter{graybox}
\label{box: Sampling Upper Bound}
\begin{svgraybox}
\textbf{Statement~\thegraybox:} When sampling, using Algorithm~\ref{alg:sampling_proc}, a circular area centered at the point \((r,\phi)\) with a radius \(c_{\kappa \%}\) and a localization resolution \(\delta \%\), the number of radial samples is upper bounded by the quantity \(S_{r,{\rm UB}}\leq \eta_{\kappa , \delta } + \frac{r a^2_{\kappa}}{2 r_{\rm RD}}(\eta_{\kappa , \delta } -1 ) + 1\) where \(\eta_{\kappa, \delta} \triangleq {a^2_{\kappa}}/{a^2_{\delta}}\) and \(a_{\kappa}\) as well as \(a_{\delta}\) are defined below Statement~\ref{box: Dr beamfocusing}.
\end{svgraybox}

Interestingly, it can be shown experimentally that the bound \(S_{r,{\rm UB}}\) becomes tighter for \(r \ll \frac{r_{\rm RD}}{a^2_{\kappa}}\), since the sampling steps are nearly uniform in that regime (i.e., \(\Delta^{\pm}_{\delta} (r) \approx \Delta^{\pm}_{\delta} (r\pm c_{\kappa \%})\)). In this case, it holds \(S_{r,{\rm UB}} \approx \eta_{\kappa , \delta } + 1\). On the other hand, for \(r\geq \frac{r_{\rm RD}}{a^2_{\kappa}}\), the sampling resolution is significantly non-uniform, and consequently, \(S_{r,{\rm UB}}\) becomes quite loose compared to the actual values of \(S_r\), which are again approximately given as \(S_r \approx \eta_{\kappa,\delta} + 1\). Finally, for an arbitrary radius \(\hat{c}\), one can find the closest \(c_{\kappa \%}\) to it so that \(\hat{c} = (1+e_c) c_{\kappa \%}\), and as a rule of thumb, \(S_{r,e_c,{\rm UB}} \triangleq  (1+e_c) S_{r,{\rm UB}}\) can be used. In the simulation Section~\ref{sec:Results}, the progression of \(S_{r,{\rm UB}}\) and the actual number of samples required with respect to increasing BS-UE distances \(r_0\) is given in Table~\ref{tbl:number_of_samples}, where it becomes apparent that not only the number of samples needed does not increase with \(r_0\), but instead it is slightly reduced.

Finally, state-of-the-art methods for tracking UE trajectories involve learning the transition density function of the motion dynamics using Particle or Kalman filtering approaches \cite{NF_tracking}\cite{RIS_and_NF_tracking}. However, such methods typically entail high computational complexity, often scaling with \(N^3\), which becomes prohibitive in the context of XL antenna arrays considered in this work. Moreover, they rely on prior assumptions, such as a constant state transition model for the dynamic variables \([r_t, \phi_t, u_t]\). Nevertheless, if one were to adopt such assumptions, the Importance Sampling mechanism within Particle Filters, as presented in \cite{NF_tracking}, could be integrated into our proposed framework. Specifically, the gridding procedure of Algorithm~\ref{alg:sampling_proc} could be employed to discretize the search area, while samples could then be drawn from the transition distribution, resulting in denser sampling in the direction of likely UE movement and sparser sampling elsewhere, thereby further reducing the total number of required samples. Such an integration is non-trivial and is left for future investigation. Overall, this chapter built upon the theoretical near-field beamforming analysis to develop a low-complexity framework for UE tracking that ensures stabilized beamforming gain performance along random trajectories, maintaining it above a predefined \(\kappa\%\) threshold. To achieve this, the concept of beam coherence time \(T_{{\rm c},\kappa\%}\) was introduced to enable dynamically adjustable estimation intervals, alongside a non-uniform sampling strategy that leverages prior position estimates to ensure a minimum of \(\delta\%\) of the optimal beamforming gain at each estimation step.

\begin{algorithm}[t]
\caption{Near-Field Beam Tracking Design}\label{alg:tracking}
  \textbf{Input:} Localization resolution \(\delta\%\), percentage of optimum beamforming gain \(\kappa\%\), previous estimates \(\hat{u}_{0},\ldots,\hat{u}_{t-1}\), \(\mathbf{\hat{p}}_{t-1}\), and \(T_{{\rm c},\kappa\%}(t-1)\), $e_c$ and $e_u$, and total number of pilots \(N_{\rm p}\).
\begin{algorithmic}[1]
\State The time interval \(T_{{\rm c},\kappa\%}(t-1)\) from the previous $(t-1)$-th estimation slot elapsed, request UE pilots' transmission.
\State Set the radius of the UE area of interest as $\hat{c}\!=\!c_{\kappa \%}(1+e_c)$.
\State Run \textbf{Alg. 1} using $\hat{r}_{t-1}$, $\hat{\phi}_{t-1}$, $\hat{c}$, and $\delta\%$, to obtain \(\left(r_{s_r},[\boldsymbol{\phi}_{s_r}]_i\right)\,\forall s_r=1,2,\dots,S_r,\) and \(\forall i=1,2,\dots, S_{\boldsymbol{\phi}_{s_r}}\). 
\State Perform the initializations \(\mathbf{g}=\mathbf{0}_{S_r\times1}\) and \(\mathbf{K}=\mathbf{0}_{S_r \times 2}\).
\State Set the averaging per \(\mathbf{\bar{Q}}\) configuration to \(M = \floor{N_{\rm p}/S_r}\).
\For{$s_r = 1,2,\ldots,S_r$} 
    \State Obtain $\mathbf{\bar{Q}}_{r_{s_r}}$ via \eqref{eq:Q_r} to steer on range \(r_{s_r}\). 
    \State Collect the measurements $\mathbf{y}_{b,r_{s_r}}[1],\ldots,\mathbf{y}_{b,r_{s_r}}[M]$ as 
    
    \hspace{-0.29cm} in~\eqref{eq:DMA_UE} for different noise realizations.
    \State Compute the average $\bar{\mathbf{y}}_{b,r_{s_r}}=M^{-1}\sum_{m=1}^{M}\mathbf{y}_{b,r_{s_r}}[m]$.
    \State Calculate \(i_{\max} = \arg\max_i |\mathbf{v}^{\rm H}_{[\bm{\phi}_{s_{r}}]_i}\mathbf{\bar{y}}_{b,r_{s_r}}|^2\) and set 
    
    \hspace{-0.29cm} $[\mathbf{g}]_{s_r} \!=\! |\mathbf{v}^{\rm H}_{[\bm{\phi}_{s_{r}}]_{i_{\max}}}\mathbf{\bar{y}}_{b,r_{s_r}}|^2$ and \([\mathbf{K}]_{s_r,:}\! =\! \left[r_{s_r},[\bm{\phi}_{s_{r}}]_{i_{\max}}\right]\).
    
\EndFor
\State Set \(s_{{r}_{\max}}\! = \!\arg \max_{i} [\mathbf{g}]_{i}\) and \(\mathbf{\hat{p}}_t=[\hat{r}_t,\hat{\phi}_t] \! = \! [\mathbf{K}]_{s_{r_{\max}},:}\).
\State Update the velocity estimate as \(\hat{u}_{t} = \frac{||\mathbf{\hat{p}}_{t} - \mathbf{\hat{p}}_{t-1}||_2}{T_{{\rm c},\kappa \%}(t-1)} \).
\State Compute the velocity estimate \(\hat{u}_{t+1}\) using \eqref{eq:predict_velocity} and the previous estimates \(\left(\hat{u}_{0},\hat{u}_{1},\ldots,\hat{u}_{t}\right)\).
\State \(c_{\kappa \%}\!=\!\min\!\left\{\!|2\hat{r}_t\sin\left(0.5\Delta_{\kappa}(\hat{\phi}_t) \right)\!|,{\!\Delta^{-}_{\kappa}\!(\hat{r}_t) }\!\right\}\) via \eqref{eq:Dr},~\eqref{eq:Dphi}. 
\State Compute the effective beam coherence
time for triggering the $(t+1)$-th estimation slot as \(T_{{\rm c},\kappa\%}(t) \!= \!\frac{c_{\kappa \%}}{\hat{u}_{t+1}(\!1+e_u\!)}\). 
\end{algorithmic}
\textbf{Output:} \(\mathbf{\hat{p}}_t\), \(\hat{u}_t\), and \(T_{{\rm c},\kappa\%}(t)\).
\end{algorithm}

\section{Simulation Results and Discussion}\label{sec:Results}

In this section, simulation results are presented to evaluate the performance of the proposed sensing-aided near-field beam tracking framework. Random Bézier trajectories were generated to model the movement of the UE, which are commonly adopted for smooth and realistic motion modeling~\cite{Graphics_Book}. Each simulation scenario involved an average over $1000$ distinct trajectories, each consisting of $100$ time steps and $6$ control points, following the procedure in \cite[Alg.~11.4.1]{Graphics_Book}. The time unit for each trajectory was configured to yield an average UE velocity of \(10\,\)m/s. Two initial UE positions were assumed to be known with perfect accuracy before executing Algorithm~\ref{alg:tracking}, and a single scatterer was placed uniformly at random within the circular area of interest surrounding the UE at each estimation phase. In the simulations, as \(r_0\) increased, the corresponding path loss, defined as \({\rm PL} \triangleq \left(\lambda/(4 \pi r_0)\right)^2\), also increased. For instance, when \(r_0 \in [5, 45]\, {\rm m}\), the SNR defined as \(P_u {\rm PL}/\sigma^2\) ranged from \(23\) to \(4\) dB, assuming a transmit power \(P_u = 5\,\)dBm and noise power \(\sigma^2 = -94\,\)dBm.

Unless otherwise specified, the following simulation parameters were used: a DMA with \(N_e = 200\) elements and \(N_m = 10\) microstrips, resulting in \(N = 2000\) metamaterials, operating at \(f_c = 30\,\)GHz with spacings \(d_e = d_m = \lambda/2\), where \(\lambda = 1\,\)cm, and a height difference \(z_0 = 1\,\)m. For the tracking parameters of Algorithm~\ref{alg:tracking}, \(N_p = 200\) pilot symbols were used, the QoS beamforming gain ratio was set to \(\kappa = 50\), and the desired estimation accuracy was \(\delta = 99\). For the auxiliary velocity-related parameters, the geometrical factor was set to \(\gamma = 2\), the velocity threshold to \(u_{\rm th} = 2.5\,\)m/s, and the error margins to \(e_c = 1.5\) and \(e_u = 0.5\).

\begin{figure}[t]
    \centering
    \includegraphics[width=\textwidth]{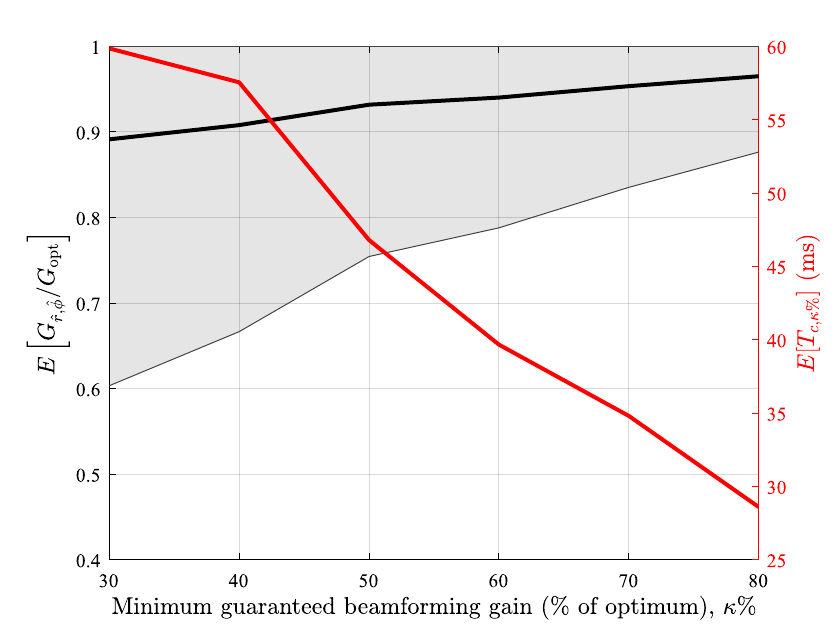}
    \vspace{-2 mm}\caption{{ The left vertical axis in black depicts the average relative beamforming gain \(E[G_{\hat{r},\hat{\phi}}/G_{\rm opt}]\) (black line) and its \(95\)-th percentile error bars (black shadowed area) over time with a time step of \(500\, {\rm \mu s}\). The right vertical axis in red includes the average estimation time interval \(E[T_{c,\kappa \%}]\) (red line) over all estimation slots.  Both are shown as functions of the $\kappa\%$ parameter.}}\vspace{-2 mm}
    \label{fig:mean_snr_over_kappa}
\end{figure}

Figure~\ref{fig:mean_snr_over_kappa} illustrates the beamforming gain performance achieved by the proposed framework under different QoS constraints, characterized by the parameter \(\kappa\%\). The left vertical axis shows the time-averaged relative beamforming gain \(E[G_{\hat{r},\hat{\phi}}/G_{\rm opt}]\), along with the area where the \(95\%\) of the data lies, represented as shadowed areas for each \(\kappa\) setting. The right vertical axis depicts the corresponding average values of \(T_{{\rm c}, \kappa \%}\) across all estimation intervals. The results verify that the achieved relative beamforming gain consistently exceeds the \(\kappa\%\) threshold with \(95\%\) confidence. In addition, average values exceed \(90\%\) even for smaller \(\kappa\%\) requirements. As expected, the mean estimation interval \(E[T_{{\rm c}, \kappa \%}]\) decreases with increasing \(\kappa\%\), reflecting the need for more frequent estimations to meet stricter QoS constraints.

\begin{figure}[t]
    \centering
    \includegraphics[width=\textwidth]{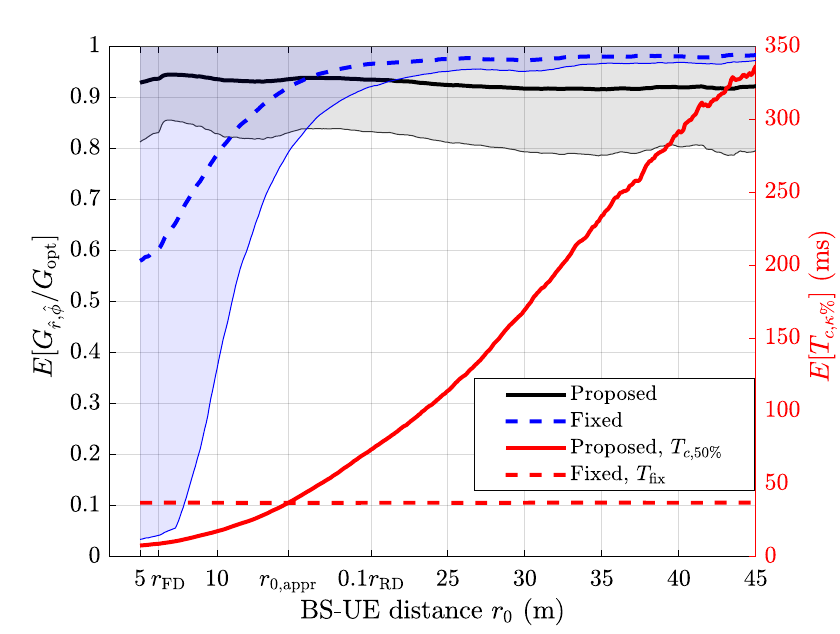}
    \caption{Average relative beamforming gain \(E[G_{\hat{r},\hat{\phi}}/G_{\rm opt}]\) and its \(90\)-th percentile error bars over time with a time step of \(500\, {\rm \mu s}\) (left vertical axis). Solid black and dashed blue lines correspond respectively to the proposed tracking algorithm and a benchmark tracking scheme with the fixed parameters \(T_{\rm fix}\triangleq E[T_{{\rm c},50 \%}]\), \(\mathcal{D} r_{\rm fix}\triangleq E[\Delta^{\pm}_{99}(r)]\), and \(\mathcal{D} \phi_{\rm fix}\triangleq E[\Delta_{99} (\phi)]\). The right vertical axis in red includes the values of the average estimation intervals for both schemes. All metrics are plotted versus the BS-UE distance \(r_0\) for $\kappa=50$.}\vspace{-2 mm}
    \label{fig:fixed_comparison_avg_gain}
\end{figure}

In Fig.~\ref{fig:fixed_comparison_avg_gain}, the performance of the proposed tracking scheme is compared against a benchmark approach that employs a fixed estimation interval \(T_{\rm fix} \triangleq E[T_{{\rm c},50 \%}]\) and fixed sampling resolutions \(\mathcal{D}r_{\rm fix} \triangleq E[\Delta^{\pm}_{99}(r)]\) and \(\mathcal{D} \phi_{\rm fix} \triangleq E[\Delta_{99} (\phi)]\), where the expectations are computed over all estimation slots of the proposed method. The left vertical axis in Fig.~\ref{fig:fixed_comparison_avg_gain} depicts the time-averaged relative beamforming gain \(E[G_{\hat{r},\hat{\phi}}/G_{\rm opt}]\) along with its \(90\)-th percentile error bars, as a function of the BS-UE distance \(r_0\), while the right vertical axis shows the average estimation intervals. As expected, the average duration between estimations increases with distance \(r_0\). Notably, both schemes achieve beamforming gains exceeding the \(50\%\) QoS threshold across the entire simulated range, although the proposed approach consistently maintains gains above \(90\%\) of the optimum. For the benchmark scheme, this gain stabilizes only after \(r_0 \geq 0.1 r_{\rm RD}\) and exhibits significant variability for smaller distances, underscoring the advantage of dynamically reconfigurable estimation intervals. As indicated by the solid red curve, the proposed scheme performs denser estimations for smaller \(r_0\) values and adapts to sparser estimations as \(r_0\) increases. For instance, at \(r_0 = 40\,\)m, the benchmark method executes approximately seven times more estimation steps than the proposed framework, while achieving a slightly higher gain of \(98\%\) compared to \(92\%\), a minimal performance difference despite the substantial reduction in estimation overhead.

To quantify the complexity reduction offered by the proposed sampling strategy, Table~\ref{tbl:number_of_samples} presents the number of samples \(S_r S_{\phi}\) generated by Algorithm~\ref{alg:sampling_proc}, alongside the number of samples \(S^{\rm HR}_r S^{\rm HR}_{\phi}\) that would be required to achieve equivalent performance using a non-dynamic sampling approach. In the latter case, the coordinate grid constructed during each estimation phase would employ fixed range and azimuth resolutions across all sampling points. For each value of \(r_0\) listed in Table~\ref{tbl:number_of_samples}, Algorithm~\ref{alg:sampling_proc} was executed with inputs \(\hat{c} = c_{50\%}\), \(\mathbf{\hat{p}}_{t-1} = [r_0, \hat{\phi}]\), and \(\delta = 99\%\), and results were averaged over all \(\hat{\phi} \in [0, \pi]\). In addition, the upper bound \(S_{r,{\rm UB}}\) defined in Statement~\ref{box: Sampling Upper Bound} is reported, along with the actual values of \(S_r\), thereby confirming the validity of the theoretical analysis. It is demonstrated that the number of samples along the range dimension remains nearly constant and always below the theoretical limit \(S_{r,{\rm UB}}\). This observation indicates that the sampling overhead introduced by the proposed near-field beam tracking strategy, relative to far-field tracking, is both upper bounded and effectively constant across a wide range of UE positions, despite the fact that the search region expands with increasing \(r_0\).

\begin{table}[t]
{ 
    \begin{center}
        \caption{Number of sampling points versus the BS-UE distance \(r_0\).}\label{tbl:number_of_samples}
        
        \resizebox{0.9\textwidth}{!}{%
            \begin{tabular}{|c| c| c| c |c| c|}
                \toprule
                \( r_0 \) (m) & \( S_r \) & \( S_r^{\text{HR}} \) & \( S_{r,\text{UB}} \) & \( S_r S_{\phi} \) & \( S_{r}^{\text{HR}} S_{\phi}^{\text{HR}} \) \\
                \hline
                \hline
                5   & 9.00     & 9.00       & 9.08 & 15.53 & 27.00 \\
                \hline
                15  & 9.00     & 76.00      & 9.18 & 30.88 & 456.00 \\
                \hline
                25  & 9.00     & 206.00     & 9.29 & 43.77 & 1648.00 \\
                \hline
                35  & 8.00     & 388.63     & 9.39 & 49.57 & 3886.27 \\
                \hline
                45  & 7.49     & 580.75     & 9.49 & 50.52 & 6946.17 \\
                \bottomrule
            \end{tabular}%
        }
    \end{center}
}\vspace{-6mm}
\end{table}

\section{Conclusion and Future Directions}\label{sec:Conclusion}

This chapter addressed the problem of sensing-aided near-field beamforming in a high-frequency point-to-point wireless communication system comprising a DMA-based BS and a mobile single-antenna UE. A theoretical characterization of the degradation of the beamforming gain relative to its optimum due to UE coordinate mismatches was provided, overcoming prior limitations such as the dependency between the range and azimuth coordinates in the derivation of angular and depth widths. In addition, the impact of microstrip losses on the near-field beamforming characteristics of the DMA was assessed and compared against the idealized lossless case. A dynamic, non-uniform polar coordinate grid was then designed for efficient sampling of the UE's circular region of interest to be used for localization at each position estimation instance. The notion of effective beam coherence time was also introduced, defined as the minimum time period during which a given QoS degradation is tolerated, and computed adaptively at the BS during each tracking step. The performance of the proposed near-field beam tracking framework, based on triggering beam sweeping only when the expected beamforming gain drops below a predefined threshold, was extensively evaluated through simulations. The theoretical findings were validated, and the proposed method was shown to outperform baseline conventional communication systems with fixed estimation intervals and uniform sampling for UE localization.

Although the current study focused on the single-user setting, the proposed framework is inherently compatible with orthogonal multiple access schemes, such as Frequency Division Multiple Access (FDMA), when combined with frequency-selective beamforming. A promising future direction could be to explore {multi-user near-field beam tracking}, where the beam coherence time, beam depth, and width metrics are redefined under multiple simultaneously focused beams from the BS, requiring new analyses. Another valuable extension involves {trajectory-aware tracking}, in which the UE's motion is modeled probabilistically using fitted path priors or stochastic mobility models to improve tracking robustness and reduce overhead. Finally, multi-BS coordination~\cite{7509379} for joint near-field beam tracking and coverage optimization presents a compelling research direction, leveraging inter-BS information exchange to enhance UE localization accuracy, beam alignment, and handover decisions across overlapping near-field regions.

\begin{acknowledgement}
This work has been supported by the Smart Networks and Services
Joint Undertaking (SNS JU) projects TERRAMETA and 6G-DISAC under
the European Union’s Horizon Europe research and innovation programme
under Grant Agreement No 101097101 and No 101139130, respectively.
TERRAMETA also includes top-up funding by UK Research and Innovation
(UKRI) under the UK government’s Horizon Europe funding guarantee.
\end{acknowledgement}

\eject

\input{editor/references}

%% file: editor/references.tex
%
%
%
%

\bibliographystyle{editor/spmpsci_unsort}

\bibliography{references}